\documentclass[onecolumn,showpacs,floatfix,superscriptaddress,amsmath,amssymb,prl]{revtex4}
\usepackage{mathrsfs}
\usepackage{amssymb}
\usepackage{graphicx}
\usepackage{hyperref}
\usepackage{tabularx}
\usepackage{array}
\usepackage{color}

\begin{document}

\title{Spin operators and representations of the Poincar\'e group}

\author{Taeseung Choi}
 \email{tschoi@swu.ac.kr}
\affiliation{Institute of General Education, Seoul Women's University, Seoul 139-774, Korea}
\affiliation{School of Computational Sciences, Korea Institute for Advanced Study, Seoul 130-012, Korea}
 \author{Sam Young Cho}
\affiliation{Centre for Modern Physics and Department of Physics,
 Chongqing University, Chongqing 400044, China}

\begin{abstract}
 We present the rigorous derivation of covariant spin operators from a general linear combination of the components of the Pauli-Lubanski vector. It is shown that only two spin operators satisfy the spin algebra and transform properly under the Lorentz transformation, which admit the two inequivalent finite-dimensional representations for the Lorentz generators through the complexification of the $SU(2)$ group. In case that the Poincar\'e group is extended by parity operation, the spin operator in the direct sum representation of the two inequivalent representations, called the new spin distinguished from the Dirac spin, is shown to be equivalent to axial and Hermitian spin operators for particle and antiparticle. We have shown that for spin $1/2$, the Noether conserved current for a rotation can be divided into separately conserved orbital and spin part for the new spin, unlike for the Dirac spin. This implies that the new spin not the Dirac spin provides good quantum observables. 

\end{abstract}
\pacs{ }

\maketitle

\section{Introduction}
\label{sec:intro}

Spin of a massive particle (e.g., electron) has become a very familiar
 and indispensable physical quantity
 in fundamental physics and applied sciences as well as quantum technologies \cite{Milestone}, whose crucial roles have been revealed in various quantum phenomena
 such as Kondo effects  \cite{Gordon},
 spin Hall effects \cite{Kato}, quantum spin fluid \cite{Xu},
 spin Hall insulator \cite{Konig},
 quantum entanglements \cite{Hasegawa} and so on. In spite of such remarkable progresses, 
there is still controversy about which spin operator is a proper spin operator even for a free particle in relativistic case \cite{Birula,Barnett,BirulaComment,BarnettReply,Bliokh17}.

Indeed, since the birth of Dirac theory in 1928 \cite{Dirac},
 many proposals have been made for a relativistic spin
 \cite{Pryce48,NW,FW,Fradkin,Chakrabarti,Hilgevoord,Fleming64,Gursey,Bogolubov,Ryder,Polyzou,Choi13,Choi15,Deriglazov,Lorce}. 
Until now, at least seven spin operators are suggested \cite{Bauke}. 
Mainly, spin operators were obtained through the construction of relativistic position operators and then the decomposition of total angular momentum into an orbital and a spin angular momentum \cite{Pryce48,NW,FW,Hilgevoord,Fleming64}. 
Some spin operators were defined by using the canonical transformations of the Dirac Hamiltonian \cite{FW} and the covariant Dirac equation \cite{Chakrabarti}, and the boost transformation of the spin in the rest frame \cite{Gursey}. 

 However, if the spin is the fundamental kinematic property of elementary particles, it should be 
determined from the space-time symmetry. It is believed that Poincar\'e symmetry is the symmetry of 
our 3+1 dimensional free space-time. 
And there are two Casimir operators in Poincar\'e group, i.e., mass and spin \cite{Bogolubov}. 
Hence spin operator should be defined by using the generators of Poincar\'e group, which are translation and Lorentz transformation operators. 

 It is well-known that the square of the Pauli-Lubanski (PL) vector is the second Casimir of Poincar\'e group, 
which is proportional to the square of spin. 
Actually the spatial components of the PL vector in the rest frame (RF) are proportional to the 
generators of the $\mathfrak{su}(2)$, i.e., $\sigma^k/2$ in usual notation, where $\sigma^k$ becomes Pauli matrix for spin $1/2$ \cite{Bogolubov}. 
However, as is known, the PL vector itself cannot be a spin operator because 
the spatial components of the PL vector in the moving frame do not satisfy 
even the basic requirement of a spin operator, i.e., the $\mathfrak{su}(2)$ algebra. 
This fact implies that the spin operator should be a linear combination of the 4-components of the PL vector, whose square 
becomes proportional to the second Casimir of Poincar\'e group. 
In fact, Bogolubov {\it et al.} \cite{Bogolubov} derived a spin operator which is a linear combination of the 
PL operator by imposing the five conditions; i) Spin operator commutes with momentum operator. 
ii) Spin operator is an axial vector. 
iii) Spin operator satisfies $su(2)$ algebra. iv) The three components of spin operator transform as the three components of 
the three-dimensional spatial vector under spatial rotation. 
v) The three components of spin operator reduce to the three spatial components of the PL vector in the RF. 
The spin operator, they derived, is unique, however, nothing but the PL vector at the RF, i.e., $\sigma^k/2$, which 
is expressed by the inverse Lorentz transformation of the PL vector in the moving frame. 
Hence it still remains an open problem to derive a relativistic spin operator other than the PL vector at the RF 
from a general linear combination of the PL vector.

 In this work we present the rigorous derivation of the two spin operators other than the PL vector at the RF,
 beginning with a general linear combination of 4-components of the PL vector by imposing the minimal physical requirements including the condition that spin should be covariant under Lorentz transformation. The covariance is required because an orbital and total angular momentum is manifestly covariant under Lorentz transformation. 
We do not require the axial vector condition. 
This condition is related with parity operation (spatial inversion), which is not necessary to find a spin operator in Poincar\'e group. 
 We find that the two spin operators are responsible for handedness (chirality) and provide corresponding two inequivalent 
spin state representations of the complexified $\mathfrak{su(2)}$ algebra, which is 
isomorphic to $sl(2,\mathbb{C})$ algebra.  
 Then the irreducible representations of the Poincar\'e group are given by momentum $p^\mu$ and spin $s$ of either the left-handed or the right-handed complexified $\mathfrak{su(2)}$ algebra. 
We will follow the usual denotation of the left-handed representation by $(s,0)$ and the right-handed representation 
by $(0,s)$, respectively, for convenience. 
The left-handed and the right-handed representations correspond to the self- and complex conjugate self-representations for spin $1/2$, respectively.  

When the Poincar\'e symmetry is extended by parity, 
the unique representation of elementary fields with spin $s$ without any redundant 
spin fields is the direct sum $(s,0)\oplus (0,s)$ representation. 
 For spin $1/2$ massive fields, the parity operator can be represented by a nontrivial linear form of spin and momentum operators. Then, as a natural consequence of parity operation for spin $1/2$ fields, 
a fundamental dynamical equation including the spin operator at the RF is obtained and found to be equivalent to the covariant Dirac equation. 
 This shows manifestly how spin presents itself in the dynamics of fields
 and why the Dirac equation successfully describes spin-$1/2$ elementary fields
 and their spin magnetic moments.

Interestingly, the two spin operators representing the left-handed and the right-handed
spin states are neither axial nor Hermitian. As a result, the spin operator, defined just by the direct sum of the left-handed and 
the right-handed spin operators in the $(s,0)\oplus (0,s)$ representation, is neither axial nor Hermitian. 
However, the Noether conserved spin angular momentum, defined by the action of the spin operator on a field as 
a state vector of the representation space, is an axial vector. 
Moreover, the Noether conserved spin angular momentum also becomes Hermitian, i.e., good observable for a particle and an antiparticle. 
We also show the conservation of spin and spin current from Noether theorem.

This paper is organized as follows. In sec. {\ref{sec:Deri}}, we derive spin operators starting with a general linear combination of the components of the PL vector operator, whose square is the second Casimir invariant of the Poincar\'e group. In sec. {\ref{sec:Repr}}, we represent the Poincar\'e group by using the two derived spin operators. The two spin operators are shown to be responsible for handedness (chirality) and to provide two inequivalent 
representations of the Poincar\'e group through the representations of the complexified $\mathfrak{su(2)}$ algebra that is isomorphic to the $sl(2,\mathbb{C})$ algebra. In sec. {\ref{sec:Fundmt}}, we extend the Poincar\'e symmetry by the parity (space inversion) and show that the parity operation on massive elementary spin $1/2$ non-chiral fields brings out the dynamical description of the fields and thus gives a fundamental dynamical equation, which is equal to the covariant free Dirac equation. 
We show in sec. {\ref{sec:Nthr}} that the spin operator defined by the action of the spin operator on an elementary field in the direct sum representation becomes an axial vector and is also manifested to become Hermitian, i.e., good observable, when it acts on either a particle or an antiparticle state.  
As an observable and conserved quantity, spin is discussed and its equivalent form is given in the particle and the antiparticle representation space. 
In section {\ref{sec:Conclusion}} our conclusions are summarized and discussed. 
In the Appendix, we relegated mathematical details on formulas in main discussions.

\section{Derivation of spin operators of the Poincar\'e group.}

\label{sec:Deri}
%
  In 1939, Wigner classified elementary particles
  by irreducible unitary representations
  of the Poincar\'e group \cite{Wigner}.
 Massive elementary particles with arbitrary spin are then considered
 as unitary irreducible representations of the Poincar\'e group.
  Yet, in the modern paradigm of elementary particles, which is quantum
  field theory,
  fields are described in general as non-unitary representations \cite{Weinberg}. 
	Straightforwardly, the normalization of the Dirac spinor, $\psi_{+\epsilon}^\dagger(p^\mu,\lambda) \psi_{+\epsilon}(p^\mu,\lambda')=\psi_{-\epsilon}^\dagger(p^\mu,\lambda) \psi_{-\epsilon}(p^\mu,\lambda')=(E/m) \delta_{\lambda \lambda'}$ , shows that the boost transformation is not unitary, where $\delta_{\lambda \lambda'}$ is Kronecker delta, $\psi_{\pm \epsilon}(p^\mu,\lambda)$ are the positive- and the negative-frequency solutions of the free Dirac equation with the spin index $\lambda$ and momentum $p^\mu=(E, {\bf p})$, and $\psi_{\pm\epsilon}^\dagger(p^\mu,\lambda)$ is the Hermitian adjoint of the $\psi_{\pm\epsilon}(p^\mu,\lambda)$.  
%
  Such a discrepancy of the unitarity might be closely connected with the definition of the spin operator, and thus requires  systematic reinvestigation of
 the irreducible representations of the Poincar\'e group to embed elementary particles into fields. 
%

\subsection{ A general form of spin operator}

One of the rigorous way to represent a group is to use the Casimir operators that
 commute with all generators of the group because the eigenvalues of the Casimir operators 
classify irreducible representations \cite{Cornwell}. 
As is well known \cite{Bogolubov},
 the Poincar\'e group has the two invariant Casimir operators giving the mass $m$ and the spin $s$ of the representation state, respectively,
 i.e.,
\begin{eqnarray}
 P^\mu P_\mu \mbox{ with the eigenvalue $m^2$}
\end{eqnarray}
 and 
\begin{eqnarray}
\label{eq:PLSQN}
W^\mu W_\mu \mbox{ with
 the eigenvalue $-m^2 s(s+1)$},
\end{eqnarray}
 where the PL vector operator $W^\mu$ is defined as
\begin{eqnarray}
 W^\mu = \frac{1}{2} \epsilon^{\,\mu\nu\rho\sigma} J_{\nu\rho} P_\sigma
\end{eqnarray}
 with the $4$-dimensional Levi-Civita $\epsilon_{\,\mu\nu\rho\sigma}$
 (we set $\epsilon_{0123}=\epsilon^{1230}=1$),
 the generators of the (homogeneous) Lorentz group $J^{\mu\nu}$,
 and the generators of translations $P^\mu$. The metric tensor $g_{\mu\nu}= \mathrm{diag} (+,-,-,-)$ will be used.
 Here, Einstein summation convention is used for the Greek indexes $\mu \in \{0,1,2,3\}$
 and will also be used for Latin indexes $k\in \{1,2,3\}$,
 unless otherwise specifically stated.
 We omit the word `operator' freely, e.g.,
 the PL vector instead of the PL vector operator, because
 the context would clarify the usage and we use capital letters for operators and small letters for usual vectors.
%
 %
%
 As is known \cite{Bogolubov}, spin operators are not identified by the PL vector itself, although the square of the PL vector reveals spin in Eq. (\ref{eq:PLSQN}), because the spatial components of the PL vector do not satisfy the $\mathfrak{su(2)}$ commutation relation, i.e., 
\begin{eqnarray}
[W^i, W^j]\neq i\epsilon_{ijk}W^k,
\end{eqnarray}
where $\epsilon_{ijk}$ is the three-dimensional Levi-Civita with $\epsilon_{123}=1$. 

However, the second Casimir in Eq. (\ref{eq:PLSQN}) allows us to consider 
the three components of a spin operator satisfying the $\mathfrak{su}(2)$ algebra 
as a linear combination of the 4-components of the PL vector such that $S^k S_k = -W^\mu W_\mu/m^2$. 
Here $S^k$ is the $k$-component of the spin 3-vector, ${\bf S}=\{ S^1, S^2, S^3 \}$. 
A general linear form of $S^k$ can be written as
 \begin{equation}
  S^k = a_{k,\mu}\ W^\mu=a_{k,0} W^0 + a_{k,k} W^k + a_{k,m\neq k} W^m,
  \label{eq:linear}
 \end{equation}
 where the index $k$ in $a_{k,k}$ term is not considered as repeated and the subscript $m\neq k$ means $m$ which is not $k$. 
%
The $SU(2)$ group generated by the $S^k$ is expected to be the little group of the Poincar\'e group, which labels a spin index.
 This requires that $S^k$ commutes with the components of momentum operator, $P^\mu$.
 Then the coefficients $a_{k,\mu}$ should be functions of complex numbers
 and $P^\mu$
 but not functions of the Lorentz generators $J_{\mu\nu}$. Note that $[P^\mu, W^\nu]=0$.
\subsection{The minimum requirements for spin operators}

Before introducing the minimum adequate requirements to determine an explicit expression of the spin operator in Eq. (\ref{eq:linear}), we need to discuss spin operator as a form of tensorial operator in order to understand clearly a transformation property of proper spin operator under Lorentz transformations. 
Similar to the total angular momentum, in 3+1 dimension, a spin angular momentum can be described by a second rank antisymmetric tensor. 
The (Hodge) dual spin tenor $\ast{S}^{\mu\nu}$ is defined as 
\begin{eqnarray}
\ast{S}^{\mu\nu} = \frac{1}{2}\epsilon^{\mu\nu\rho\sigma}S_{\rho\sigma}, 
	\end{eqnarray}
	by the spin tensor $S_{\rho\sigma}$ and then the $S^k$ can be mapped from the $k0$-component of the dual spin tensor
\begin{eqnarray}
\label{eq:k0SP}
 \ast{S}^{k0}  =\frac{1}{2} \epsilon_{kij}S^{ij}=S^k,
\end{eqnarray}
where $\epsilon^{k0ij}$ becomes the three-dimensional Levi-Civita $\epsilon_{kij}$. 
The mapping in Eq. (\ref{eq:k0SP}) shows that the $k$-component of the spin ${\bf S}$ should transform as a $k0$-component of second rank tensor under Lorentz	boost, not just the $k$-component of a spatial three-vector. 

 Actually, we find the minimum requirements determining the coefficients $a_{k,\mu}$ in Eq. (\ref{eq:linear}): 

$\bullet$ $S^k$ should commute with the momentum; 
\begin{eqnarray}
\label{eq:LGC}
[S^k, P^\mu]=0 \mbox{ (little group condition).}
\end{eqnarray}

 $\bullet$ $S^k$ should satisfy the $\mathfrak{su}(2)$ algebra; 
\begin{eqnarray}
\label{eq:AMC}
[S^i, S^j]=i\epsilon_{ijk} S^k \mbox{ (angular momentum condition).}
\end{eqnarray}

 $\bullet$  $S^k$ should transform as $k0$-component of a second-rank tensor under Lorentz transformation;
\begin{eqnarray}
\label{eq:TNSC}
  i[J^{\mu\nu}, \ast{S}^{k0}]=g^{\nu k} {\ast{S}^{\mu 0}} - g^{\mu k} {\ast{S}^{\nu 0}} - g^{0 \mu } {\ast{S}^{k \nu}} + g^{0 \nu } {\ast{S}^{k \mu}} \mbox{ (tensor condition).}
\end{eqnarray} 
The tensor condition guarantees the spin to be covariant from a construction. 

\subsection{Brief explanation of the derivation of spin operators}
In order to derive spin operators, one can implement the requirements on the general expression of spin operators in Eq. (\ref{eq:linear}). The little group condition (\ref{eq:LGC}) was used to let the coefficients $a_{k,\mu}$ be functions of complex numbers and momentum. As the sub-condition of the tensor condition, the three-dimensional vector condition 
\begin{eqnarray}
\label{eq:VCD}
[J^j, S^k]=i\epsilon_{jkl}S^l,
\end{eqnarray}
fixes the functional form of the coefficients $a_{k,\mu}$ that gives (Appendix A (i))
\begin{eqnarray}
 S^k = f_4(P^0) P^k W^0 + f_1(P^0) W^k + \ f_3(P^0)\ \epsilon_{kml}\ P^l W^m,
 \label{eq:sol1}
 \end{eqnarray}
where $f_1(P^0)$, $f_3(P^0)$, and $f_4(P^0)$ are functions of $P^0$.

Next we impose the angular momentum condition (\ref{eq:AMC}) on the spin $S^k$ in Eq. (\ref{eq:sol1}). The angular momentum condition gives the relations among $f_1(P^0)$, $f_3(P^0)$, and $f_4(P^0)$ as 
\begin{subequations}
 \begin{eqnarray}
 f_4 &=& - f_4\ f_1\ P^0 - f_1^2 +\ m^2\ f_3^2 ,
 \label{eq:ef1}
 \\
 f_1 &=& f_4\ f_1\ (P_0^2 - m^2) +\ f_1^2\ P^0,
 \label{eq:ef2}
 \\
 f_3 &=& f_4\ f_3\ (P_0^2 - m^2) +\ f_1\ f_3\ P^0,
 \label{eq:ef3}
 \end{eqnarray}
\end{subequations}
where $P^2_0\equiv(P^0)^2$ and the argument $P^0$ of the functions $f_1$, $f_3$, and $f_4$ are omitted for simplicity. 
Interestingly, these relations among the three unknown function $f$s show that if $f_3=0$, the functions $f_1$ and $f_4$ are specified explicitly by using the relations (\ref{eq:ef1}) and (\ref{eq:ef2}). The $f_3$ term in Eq. (\ref{eq:sol1}) is non-axial and thus the spin $S^k$ for $f_3=0$ becomes the same axial vector as Bogolubov {\it et al.} obtained in Ref. \cite{Bogolubov} (Appendix A (iv)). Note that for nonzero three functions $f_1$, $f_3$, and $f_4$, Eqs. (\ref{eq:ef2}) and (\ref{eq:ef3}) give the same relation, i.e., 
\begin{eqnarray}
f_4(P^2_0 -m^2)+ f_1 P^0=1,
\end{eqnarray} and then the three unknown functions $f_1$, $f_3$, and $f_4$ cannot be specified. Hence we need more condition to determine spin operator explicitly. 

One can easily find that the tensor condition, which requires every terms of $S^k$ in Eq. (\ref{eq:sol1}) should transform as a $k0$-component of second-rank tensor under Lorentz transformation, can determine the specific functional form of $f_1(P^0)$, $f_3(P^0)$, and $f_4(P^0)$ (Appendix A (iii)). For example, $f_3(P^0)$ should be $c$, where $c$ is a Lorentz-invariant function of momentum (scalar under Lorentz transformation), because $\epsilon_{kml}P^l W^m=\epsilon_{0kml}P^l W^m$ transforms as a $k0$-component of second-rank tensor. We then obtain the specific functional forms as $f_1(P^0)= b P^0$ and $f_4(P^0)=a$, where $a$ and $b$ are also Lorentz-invariant functions of momentum. 
Hence the relations among $a$, $b$, and $c$ are given as
 \begin{subequations}
 \begin{eqnarray}
 a  &=& - a\ b\ P_0^2 - b^2 P_0^2 + m^2\ c^2,
 \label{eq:f1aa}
 \\
 b &=& a\  b\ (P_0^2 - m^2) + b^2\ P_0^2,
 \label{eq:f2bb}
 \\
 c &=& a\  c\ (P_0^2 - m^2) + b\ c\ P_0^2.
 \label{eq:f3cc}
 \end{eqnarray}
\end{subequations}
For any $P^0$ related with boost transformations (Appendix A (iii)), the two solutions are given as
\begin{eqnarray}
 a  =-\frac{1}{m^2}, ~~~b=\frac{1}{m^2}, ~~~c=\pm \frac{i}{m^2}.
 \label{eq:twosol}
 \end{eqnarray}
 The corresponding two spin operators are explicitly obtained as 
 \begin{equation}
 S^k_{\pm} = \frac{1}{m^2} \left( P^0 W^k - W^0 P^k\right)
 \pm \frac{i}{m^2} \epsilon_{0klm}  W^l P^m.
 \label{eq:pm}
 \end{equation}
These two spin operators are the same as those determined by constructing antisymmetric tensor from the commutator of the Pauli-Lubanski vector, which satisfy the generator algebra of the homogeneous Lorentz group, in Ref. \cite{Ryder}. 

In the next sections, we will represent the Poincar\'e group by using these two spin operators and rederive the covariant Dirac equations for a particle and an antiparticle. It will be cleared that the parity operator plays a crucial role to derive the covariant Dirac equations for a particle and an antiparticle. This fact implies that the space-time symmetry for a free Dirac field is not just the Poincar\'e symmetry but the parity-extended Poinca\'e symmetry.   

\subsection{The corresponding spin tensor operators}

Each of the derived spin three-vectors $S^k_\pm$ in Eq. (\ref{eq:pm}) has the corresponding dual spin tensor with $S^k_\pm = \ast{S}^{k0}_\pm$.  
Using Eq. (\ref{eq:TNSC}) with Eqs. (\ref{eq:k0SP}) and (\ref{eq:pm}), there are two possible candidates for the $ij$-components of the dual spin tensor operators as follows
\begin{subequations}
\begin{eqnarray}
\ast{S}^{ij}_{\pm;1}&=&\frac{1}{m^2} (W^i P^j -P^i W^j)\pm \frac{i}{m^2} \epsilon\,^{ij\rho\sigma} W_\rho P_\sigma
\label{eq:DSP1}
\\
\ast{S}^{ij}_{\pm;2}&=&-\left[\frac{1}{m^2} (W^i P^j -P^i W^j)\pm \frac{i}{m^2} \epsilon\,^{ij\rho\sigma} W_\rho P_\sigma\right].
\label{eq:DSP2}
\end{eqnarray}
\end{subequations}
The two corresponding spin tensors defined by 
\begin{eqnarray}
S^{\mu\nu}_{\pm;n}=-\frac{1}{2}\epsilon^{\mu\nu\rho\sigma}{\ast S}_{\pm;n;\rho\sigma}
\end{eqnarray}
through ${\ast{\ast S}}^{\mu\nu}_{\pm;n}=-S^{\mu\nu}_{\pm;n}$
satisfy the same commutation relations of angular momentum. Here $n=\{1, 2\}$.  
We choose the second one in Eq. (\ref{eq:DSP2}) as the definition of $\ast S^{ij}_\pm$ because this choice allows the usual definition of the spin tensor at the RF with the commutator of gamma matrices in Eq. (\ref{eq:RFGAMMA}) and thus we will omit the subscript $2$ in the spin. 
The six components of the antisymmetric spin tensor $S^{\mu\nu}_\pm$ are not independent because of the relation
\begin{eqnarray}
S^{0k}_\pm =\pm i S^k_\pm \mbox{  and } S^{ij}_\pm= \epsilon_{kij}S^{k}_\pm.
\end{eqnarray}
Consequently, the $S^k_\pm$ determine all components of the antisymmetric spin tensor $S^{\mu\nu}_\pm$. 

%

%

Straightforwardly, one can easily see that both $S^k_\pm$ in Eq. (\ref{eq:pm}) reduce to the $k$-component of the PL vector at the RF. As for another consistency check,
 one can show that the spin operators $S^k_\pm$ give the second Casimir, i.e., $S^k_\pm S^k_\pm = - W^\mu W_\mu/m^2 $.
 Then the two spin operators offer the same Casimir operator
 of the Poincar\'e group, i.e., $S^k_+ S^k_+ = S^k_- S^k_-$. 
 %
 However, the two spin operators do not commute with each other, i.e., $[S^i_+, S^j_-]\neq 0$. 
This implies that the irreducible representations of the Poincar\'e group are not the tensor product of each representation of $S^k_+$ and $S^k_-$. This fact is different from the representation of the homogeneous Lorentz group \cite{Schwartz}, in which the generators of two $SU(2)$ group commute each other, because the homogeneous Lorentz group can be expressed as $SO(3,1) \sim SU(2) \otimes SU(2)$. Therefore, the finite-dimensional representations of the homogeneous Lorentz group are generally given by two angular momenta $(m/2, n/2)$, where $m$, $n$ are integers. On the other hand, the two spin operators $S^k_\pm$ will be shown to give two inequivalent representations even for the same spin $s=s_+=s_-$. 
 The two representations are associated with the different transformation properties under the Lorentz boost transformations, i.e., handedness (chirality).
 It will become clear in detailed discussions of the following section.

\section{Representations of the Poincar\'e group using the two spin operators}
\label{sec:Repr}

\subsection{Representation of the Poincar\'e group and handedness}
As we have shown, the Poincar\'e group has the two Casimir operators $P^\mu P_\mu$ and either $S^k_+ S^k_+$ or $S^k_- S^k_-$.
 All representation spaces of the Poincar\'e group are then classified by the eigenvalues
 of the two Casimir operators as $(m, s_\pm)$ with integer or half-integer spin $s_\pm$. 
Then the base states $\Psi_{\pm}(p^\mu,\lambda_\pm)$
 of the representation spaces $(m,s_\pm)$, on which the representations of the Poincar\'e group act, satisfy
 the following eigenvalue equations of momentum and spin operators
\begin{subequations}
\begin{eqnarray}
 P^\mu\ \Psi_{\pm}(p^\mu,\lambda_\pm) &=& p^\mu\, \Psi_{\pm}(p^\mu,\lambda_\pm),
 \label{eq:psk}
 \\
 S^k_\pm\  \Psi_{\pm}(p^\mu,\lambda_\pm) &=& \lambda_\pm \Psi_{\pm}(p^\mu,\lambda_\pm),
\label{eq:sk}
\end{eqnarray}
\end{subequations}
 where the momentum
 $p^\mu=(p^0,\mathbf{p})$ and the spin index $\lambda_\pm \in \{-s_\pm, -s_\pm+1, \cdots, s_\pm -1, s_\pm\}$. 
Hereafter we use $s=s_+=s_-$ and $\lambda=\lambda_+=\lambda_-$ for simplicity.  
%
%
%
%

The spin operators $S^k_\pm$ in Eq. (\ref{eq:pm}) are dependent on the momentum operator $P^\mu$. Hence, after acting on the base state $\Psi_{\pm}(p^\mu,\lambda)$, the spin operators $S^k_\pm$ can be denoted as the spin operator $S^k_\pm ({p}^\mu)$ for the given momentum $p^\mu=(p^0, {\bf p})$. Then let $\psi_\pm(p^\mu,\lambda)$ be the eigenstate of the spin $S^k_\pm({p^\mu})$, which becomes a spinor for half integer $s$. For a given $p^\mu$, there are $2s+1$ independent spin eigenstates. Then it is convenient to describe a momentum eigenstate separately from the base state $\Psi_\pm(p^\mu,\lambda)$ as  
\begin{eqnarray}
\label{eq:PF}
\Psi_\pm(p^\mu,\lambda)= \psi_\pm(p^\mu,\lambda) |p^\mu\rangle, 
\end{eqnarray}
where $|p^\mu\rangle$ is the ordinary eigenstate of the momentum $P^\mu$. 
The momentum operator $P^\mu$ operates on the spin eigenstate $\psi_\pm(p^\mu,\lambda)$ trivially as a c-number. That is, 
\begin{subequations}
\begin{eqnarray}
P^\mu \psi_\pm(p^\mu,\lambda)|p^\mu\rangle &=& p^\mu\psi_\pm(p^\mu,\lambda)|p^\mu\rangle, 
\label{eq:MST} \\
S^k_\pm({p^\mu})\psi_\pm(p^\mu,\lambda)&=&\lambda \psi_\pm(p^\mu,\lambda), 
\label{eq:SPST} 
\end{eqnarray}
\end{subequations}
with the detailed action steps
\begin{eqnarray}
S^k_\pm \Psi_\pm(p^\mu,\lambda)= S^k_\pm({p^\mu}) \psi_\pm(p^\mu,\lambda) |p^\mu\rangle =\lambda \psi_\pm(p^\mu,\lambda_\pm) |p^\mu\rangle.
\end{eqnarray}
Note that the product form in Eq. (\ref{eq:PF}) is not a tensor product because the spin state $ \psi_\pm(p^\mu,\lambda) $ is not determined independently from the momentum eigenstate $|p^\mu\rangle$. The $S^k_\pm({p^\mu})$ can be considered and used as the spin operators for the field with the specific momentum $p^\mu=(p^0, {\bf p})$. The explicit form of $S^k_\pm({p^\mu})$ will be given later in this section.

Let us construct two inequivalent representations of Poincar\'e group elements in the space generated by $\Psi_\pm(p^\mu,\lambda)$. 
As usual, the translation operator is represented by $e^{ip^\mu a_\mu}$ that transforms 
\begin{eqnarray}
\Psi_\pm(p^\mu, \lambda) \rightarrow e^{i p^\mu a_\mu}\Psi_\pm(p^\mu,\lambda) 
\end{eqnarray} 
with an arbitrary constant 4-vector $a_\mu$ for the translation parameter. 
The two spin operators $S^k_\pm(p^\mu)$ admit the following two inequivalent finite-dimensional representations of the homogeneous Lorentz group, which is isomorphic to the complexified ${SU}(2)$ group, acting on the space generated by $\psi_\pm(p^\mu,\lambda)$; 
\begin{eqnarray}
\label{eq:IRLT}
e^{i {\bf S}_+(p^\mu)\cdot(\mathbf{\theta}-i \mathbf{\xi})} \mbox{ and }
 e^{i {\bf S}_-(p^\mu) \cdot(\mathbf{\theta}+i \mathbf{\xi})}=\left[ \left( e^{i {\bf S}_+(p^\mu)\cdot(\mathbf{\theta}-i \mathbf{\xi})} \right)^{-1} \right]^\dagger,
\end{eqnarray}
where $(S^k_+(p^\mu))^\dagger = S^k_-(p^\mu)$ from Eq. (\ref{eq:pm}) is used, and $\mathbf{\theta}$ is a compact parameter indicating a rotation angle and $\mathbf{\xi}$ is a non-compact parameter indicating a rapidity. The Lorentz transformation with these $\mathbf{\theta}$ and $\mathbf{\xi}$ transforms the momentum $p^\mu$ to $q^\mu=\Lambda^\mu_{\phantom{\mu}\nu}(\mathbf{\theta}, \mathbf{\xi})p^\nu$.  
That is, the state $\Psi_\pm(p^\mu,\lambda)$ transforms under Lorentz transformations according to 
\begin{eqnarray}
\label{eq:LTSPN}
e^{i {\bf S}_+({p^\mu})\cdot(\mathbf{\theta}-i \mathbf{\xi})} \psi_+(p^\mu,\lambda)|q^\mu\rangle \mbox{ and }
 e^{i {\bf S}_-({p^\mu}) \cdot(\mathbf{\theta}+i \mathbf{\xi})}\psi_-(p^\mu,\lambda)|q^\mu\rangle ,
\end{eqnarray} 
respectively. $e^{i {\bf S}_+({p^\mu})\cdot(\mathbf{\theta}-i \mathbf{\xi})}$ and $e^{i {\bf S}_-({ p^\mu}) \cdot(\mathbf{\theta}+i \mathbf{\xi})}$ only change the spin index in general, hence we call these the spin state representation of the homogeneous Lorentz transformation.

Actually, $e^{i {\bf S}_+({p^\mu})\cdot(\mathbf{\theta}-i \mathbf{\xi})}$ and $e^{i {\bf S}_-({ p^\mu}) \cdot(\mathbf{\theta}+i \mathbf{\xi})}$ cannot be mapped into each other by a similarity transformation and therefore provide two inequivalent spin state representations of the Lorentz transformations in the specific frame of the state with momentum ${p}^\mu$. Note that the $S^k_\pm({p^\mu})$ are the linear combinations of the Lorentz generators $J^{\mu\nu}$, because the momentum $p^\mu$ is a value not an operator in this specific frame. 
For spin $1/2$, the two representations $e^{i {\bf S}_+({p^\mu})\cdot(\mathbf{\theta}-i \mathbf{\xi})}$ and $e^{i {\bf S}_-({p^\mu}) \cdot(\mathbf{\theta}+i \mathbf{\xi})}$ become a self-representation and equivalent to a complex conjugate self-representation in two-dimension with explicit forms in Eq. (\ref{eq:spin}), respectively \cite{Harald}. 
Usually the self-representation and the complex conjugate self-representation are called the left-handed and the right-handed spinor representation, which are denoted by $(1/2,0)$ and $(0,1/2)$ for the homogeneous Lorentz group, respectively. 
Following this conventional denotation, we will also call the spin state representations $e^{i {\bf S}_+({p^\mu})\cdot(\mathbf{\theta}-i \mathbf{\xi})}$ and $e^{i {\bf S}_-({p^\mu}) \cdot(\mathbf{\theta}+i \mathbf{\xi})}$ 
as the left-handed and the right-handed representations, respectively, and denote by $(s,0)$ and $(0, s)$. 
Consequently, the two spin operators $S^k_\pm$ provide the handedness (chirality) to the states  $\psi_{\pm}(p^\mu,\lambda)$ and $\Psi_{\pm}(p^\mu,\lambda)$.

Note that the left-handed and the right-handed representations of the Poincar\'e group do not need the trivial scalar representation with spin $0$ in $(1/2,0)$ and $(0,1/2)$ for the homogeneous Lorentz group. The trivial scalar representation for spin $0$ is $1$, so this does not give any change to the representation of spin $1/2$ mathematically, however, the physical meaning of this trivial scalar representation is not clear. This fact implies that the symmetry for a spin $1/2$ particle is at least the Poincar\'e symmetry not the homogeneous Lorentz symmetry.   

\subsection{Explicit representations of the two spin operators in an arbitrary reference frame}

It is helpful to obtain the explicit representations of the two spin operators $S^k_\pm({p^\mu})$ for the states $\Psi_\pm(p^\mu,\lambda)$ in an arbitrary reference frame with an arbitrary momentum $p^\mu$. It can be accomplished most easily by using a Lorentz transformation (LT) of $p^\mu$ and the PL vector $w^\mu$. To this purpose, it is enough to consider a pure boost transformation from the RF (so called the standard LT) because a rotation in the RF does not change the momentum of the field $\Psi_\pm(p^\mu,\lambda)$. 
Two successive non-collinear Lorentz boosts from the RF,
 equivalent to an effective rotation in the RF followed by an effective standard LT,
 are well-known to give rise to a nontrivial effect \cite{Wigner,Weinberg64,Choi14}.
 However, such an effective rotation in the RF is also not relevant to obtain
 the representation of the spin operators in the moving frame. 

 The standard LT, carrying from $k^\nu=(m,{\bf 0})$ to $p^\mu=L^\mu_{\phantom{\nu}\nu}k^\nu$, is given as 
\begin{eqnarray}
\label{eq:SLTST}
 L^0_{\ 0} = \frac{p^0}{m}, \mbox{\phantom{ }} L^i_{\ 0} = \frac{p^i}{m}, \mbox{ and } 
L^i_{j} = \delta_{ij} + \frac{p^i p^j}{m(p^0+m)}.
\end{eqnarray}
The PL vector becomes $w^\mu_{rest} = (0, m {\boldsymbol{\sigma}}/2)$ at the RF, where ${\boldsymbol{\sigma}}/2$ is the $(2s+1)$-dimensional representation of $\mathfrak{su}(2)$ algebra. 
%
%
Thus the PL vector in the moving frame 
is given as
\begin{eqnarray}
 w^0 = \frac{\boldsymbol{\sigma} \cdot \mathbf{p}}{2} 
\mbox{ and  } w^i = m \frac{\sigma^i}{2} + p^i \frac{\boldsymbol{\sigma} \cdot \mathbf{p}}{2(m+p^0)},
\end{eqnarray} 
where ${\boldsymbol{\sigma}}\cdot \mathbf{p}=\sigma^k p^k$.
 As a result, one can find the explicit expressions of the spin operators $S^k_\pm({p^\mu})$ in a moving frame as 
  \begin{equation}
  S^k_\pm ({p}^\mu)
   = \frac{p^0}{2m} \sigma^k
        - \frac{ p^k ({\boldsymbol{\sigma}} \cdot \mathbf{p})} {2m (p^0 + m)}\
         {\pm}\ i \frac{1}{2m} \left({\boldsymbol{\sigma}} \times \mathbf{p}\right)^k .
  \label{eq:spin}
  \end{equation}
This Eq. (\ref{eq:spin}) shows that the spin operator at the RF is represented by 
\begin{eqnarray}
\label{eq:SPRF}
S^k_\pm({ k}^\mu)=\frac{\sigma^k}{2},
\end{eqnarray}
i.e., the usual $(2s+1)$-dimensional matrices satisfying spin algebra.

Then the spin state representation of the standard LT becomes $
e^{\pm {\bf S}_\pm({ k}^\mu)\cdot \mathbf{\zeta}}=e^{\pm {\boldsymbol{\sigma}}\cdot \mathbf{\zeta}/2} $, will be denoted by ${\mathnormal{Q}}_\pm({\bf p})$ 
for simplicity, 
where ${\mathbf{\zeta}}= 2\, \hat{ \mathbf{p}} \tanh^{-1} [ |\mathbf{p}|/ (p^0+m) ]$. 
Then the standard LT ${\mathnormal{Q}}_\pm({\bf p})$ transforms the spin states $\psi_\pm(k^\mu,\lambda)$ at the RF to $\psi_\pm(p^\mu,\lambda)$ at the moving frame as
\begin{eqnarray}
\label{eq:TSPIN}
\psi_\pm(p^\mu,\lambda)= \mathnormal{Q}_\pm({\bf p})\psi(k^\mu,\lambda).
\end{eqnarray}
Interestingly, the explicit expression of the spin operator $S^k_\pm({p^\mu})$ in Eq. (\ref{eq:spin}) can also be obtained by the similarity transformation of the spin operator at the RF for ${\mathnormal{Q}}_\pm({\bf p})$ as (Appendix B)
\begin{equation}
  S^k_\pm ({p}^\mu) = {\mathnormal{Q}}_\pm({\bf p}) S^k_\pm({k}^\mu) {\mathnormal{Q}}_\pm^{-1}({\bf p}),
 \label{eq:covariance}
\end{equation}
where ${\mathnormal{Q}}_\pm^{-1}({\bf p})$ is the inverse of ${\mathnormal{Q}}_\pm({\bf p}) $. 
Using Eqs. (\ref{eq:TSPIN}) and (\ref{eq:covariance}), the eigenvalue equation at the RF,
\begin{eqnarray}
S^k_\pm({k}^\mu) \psi(k^\mu,\lambda)=\lambda  \psi(k^\mu,\lambda),
\end{eqnarray}
derives the equation, 
\begin{eqnarray}
S^k_\pm({p^\mu}) \psi_\pm(p^\mu,\lambda)= \lambda \psi_\pm(p^\mu,\lambda).
\end{eqnarray} 
This fact implies that the standard LT does not change the eigenvalue of the spin operator as expected by the Wigner rotation \cite{Weinberg64}. 

For the general Lorentz transformation $\Lambda$ that transforms $p^\mu$ to $q^\mu=\Lambda^\mu_{\phantom{\mu}\nu}p^\nu$, the state transforms under $U(\Lambda)$ as
\begin{eqnarray}
\label{eq:WURT}
U(\Lambda)\psi(p,\lambda) = \sqrt{\frac{(\Lambda p)^0}{p^0}} \mathcal{D}^{(s)}_{\lambda' \lambda}\left( \Omega(\Lambda,p) \right)\psi(\Lambda p, \lambda')
\end{eqnarray}
where $\Omega(\Lambda,p)$ is the Wigner rotation given by
\begin{eqnarray}
\Omega(\Lambda,p)=L^{-1}(\Lambda p) \Lambda L(p)
\end{eqnarray}
with $L(p)$ in Eq. (\ref{eq:SLTST}) \cite{Weinberg}. The representation $\mathcal{D}^{(s)}$ is unitary, however, not covariant in the following sense: The state $U(\Lambda)\psi(p,\lambda)$ can also be obtained from
\begin{eqnarray}
\sqrt{\frac{(\Lambda p)^0}{p^0}} U(\mathcal{R})\left(\Omega(\Lambda,p)\right) \psi(\Lambda p, \lambda)=\sqrt{\frac{(\Lambda p)^0}{p^0}} \mathcal{D}^{(s)}_{\lambda' \lambda}\left( \Omega(\Lambda,p) \right)\psi(\Lambda p, \lambda'),
\end{eqnarray}
where $\mathcal{R}$ is the rotation \cite{Weinberg}. That is, the representation $\mathcal{D}^{(s)}$ is not defined in the specific reference frame. 

The new representations $\exp{i {\bf S}_\pm(p^\mu)\cdot (\boldsymbol{\theta} \mp i\boldsymbol{\xi})}$ of the Lorentz generator is defined in an arbitrary frame and gives the Wigner rotation naturally as
\begin{eqnarray} \nonumber
e^{i {\bf S}_\pm(p^\mu)\cdot (\boldsymbol{\theta}\mp i \boldsymbol{\xi})} \psi_{\pm}(p,\lambda)&=& e^{\pm{\bf S}_\pm(k^\mu) \cdot \boldsymbol{\xi}'} e^{i {\bf S}_\pm(k^\mu)\cdot \boldsymbol{\Omega}(\Lambda,p)}e^{\mp{\bf S}_\pm(k^\mu) \cdot \boldsymbol{\xi}'} e^{\pm{\bf S}_\pm(k^\mu) \cdot \boldsymbol{\xi}'} \psi(k^\mu,\lambda) \\
&=& e^{i {\bf S}_\pm(q^\mu)\cdot \boldsymbol{\Omega}(\Lambda,p)/2} \psi_\pm(q^\mu,\lambda),
\end{eqnarray}
where $e^{\pm{\bf S}_\pm(k^\mu) \cdot \boldsymbol{\xi}'}$ is the standard LT that transforms from the rest frame to the frame with $q^\mu$.

\section{$(s,0)\oplus(0,s)$ representation and the fundamental dynamical equation for free spin $1/2$ massive fields}
\label{sec:Fundmt}
 One can notice that the two spin operators $S^k_\pm$ in Eq. (\ref{eq:pm}) are non-Hermitian, which implies that $S^k_\pm$ are not good spin observables. 
It will be shown that the Hermitian spin as a good observable can be obtained in the representation of the parity-extended Poincar\'e group. 
 
\subsection{Representation of the parity-extended Poincar\'e group}

In this section, we will study the parity-extended Poincar\'e group, which has parity transformation (space inversion) as well as translations and Lorentz transformations. The two inequivalent spin state representations of the Poincar\'e group play a fundamental role
 as the building blocks for the irreducible representations of the parity-extended Poincar\'e symmetry. 

One can easily notice that the two inequivalent spin state representations of the Poincar\'e group are connected each other by space inversion. Under parity, the  momentum and the PL vector
 transform as, respectively, 
\begin{eqnarray}
p^\mu=(p^0, \mathbf{p}) \leftrightarrow \tilde{p}^\mu=(p^0, -\mathbf{p}) \mbox{ and }
w^\mu=(w^0, \mathbf{w}) \leftrightarrow  \tilde{w}^\mu =(-w^0, \mathbf{w}).
\end{eqnarray}
 Accordingly, 
$S^k_+({p^\mu})$ transforms to $S^k_-({ p^\mu})$ and vice versa in Eq. (\ref{eq:pm}) under parity:
\begin{eqnarray}
S^k_+({p^\mu}) \longleftrightarrow S^k_+(\tilde{p}^\mu)=S^k_-({p^\mu}),
\end{eqnarray}
and the spin state $\psi_+(p^\mu,\lambda)$
 transforms to the $\psi_-(p^\mu,\lambda)$ and vice versa:
\begin{eqnarray}
 \psi_+(p^\mu,\lambda) \longleftrightarrow \psi_+(\tilde{p}^\mu,\lambda)=\psi_-(p^\mu,\lambda)
\end{eqnarray}
with Eq. (\ref{eq:TSPIN}) and ${\mathnormal{Q}}_\pm(-{\bf p})={\mathnormal{Q}}_\mp({\bf p})$. 
Hence the spin eigenvalue equation transforms as
\begin{eqnarray}
\label{eq:SEGP}
S^k_\pm(p^\mu)\psi_\pm(p^\mu,\lambda)=\lambda \psi_\pm(p^\mu,\lambda) \longrightarrow  S^k_\mp(p^\mu)\psi_\mp(p^\mu,\lambda)=\lambda \psi_\mp(p^\mu,\lambda)
\end{eqnarray}
with the same eigenvalue $\lambda$, which shows that the spin eigenvalue does not change under parity.
%
%
 
To construct a representation theory involving parity operation properly,
 the corresponding representation space requires both the left-handed and the right-handed representation of the Poincar\'e group. 
%
 Among the possible spin state representations of the parity-extended Poincar\'e group
 the natural representation, 
 which is the only representation without any redundant representation space, is the $(s,0)\oplus (0,s)$ representation. 
To obtain the representation of the $(s,0) \oplus (0,s)$ space, we should construct the base spin states and a natural choice of the base (spin) states is 
\begin{eqnarray}
\label{eq:PST}
\psi^P(p^\mu,\lambda) = \left( \begin{array}{c} \psi_+(p^\mu,\lambda) \\ \psi_-(p^\mu,\lambda) \end{array} \right)
\end{eqnarray} 
with $\psi_+(p^\mu,\lambda) \leftrightarrow \psi_-(p^\mu,\lambda)$ under parity. 

The $\psi^P(p^\mu,\lambda) $ gives only $2s+1$ independent states, however, the dimension of the $(s,0) \oplus (0,s)$ representation space is twice of $2s+1$. Hence it is desirable to obtain the other $2s+1$ base states that are independent on the $\psi^P(p^\mu,\lambda) $. To this purpose, we define the following Lorentz invariant scalar product as
\begin{eqnarray}
\label{eq:SPD}
\psi^{P\dagger}(p^\mu,\lambda) \mathcal{P} \psi^P(p^\mu,\lambda)=\psi^\dagger_+(p^\mu,\lambda) \psi_-(p^\mu,\lambda) + \psi^\dagger_-(p^\mu,\lambda) \psi_+(p^\mu,\lambda),
\end{eqnarray}
where $\mathcal{P}$ is the parity operator. The base state $\psi^P(p^\mu,\lambda)$ is transformed from the spin state $\psi^P(k^\mu,\lambda)$ at the RF by using the standard LT defined in the $(s,0) \oplus (0,s)$ representation similar to the way in the Poincar\'e group as 
\begin{eqnarray}
\label{eq:PLT}
 \psi^P(p^\mu,\lambda) 
 =  {\mathnormal{Q}} ({\bf p}) \psi^P(k^\mu,\lambda),
\label{eq:dstateLT}
\end{eqnarray}
where $\mathnormal{Q}({\bf p})$ is the direct sum of ${\mathnormal{Q}}_+ ({\bf p})$ and ${\mathnormal{Q}}_- ({\bf p})$, i.e., ${\mathnormal{Q}}_+ ({\bf p}) \oplus {\mathnormal{Q}}_- ({\bf p})$. 
One can check the invariance of the scalar product under the standard LT as
\begin{eqnarray}
\psi^{P\dagger}(p^\mu,\lambda) \mathcal{P} \psi^P(p^\mu,\lambda) 
&=& \psi^{P\dagger}(k^\mu,\lambda) \mathcal{P}\mathcal{P}\mathnormal{Q}^\dagger({\bf p})\mathcal{P} \mathnormal{Q}({\bf p}) \psi^P(k^\mu,\lambda) \\ \nonumber
&=& \psi^{P\dagger}(k^\mu,\lambda) \mathcal{P} \psi^P(k^\mu,\lambda) 
\end{eqnarray}
using $\mathcal{P}^2=\mathbb{I}_{2(2s+1)}$ and $\mathcal{P}\mathnormal{Q}^\dagger({\bf p})\mathcal{P}=\mathcal{P}\mathnormal{Q}({\bf p})\mathcal{P}=\mathnormal{Q}^{-1}({\bf p})$ because of $\mathnormal{Q}^{-1}({\bf p})=\mathnormal{Q}(-{\bf p})$. 
Using the scalar product in Eq. (\ref{eq:SPD}), the other $2s+1$ base states orthogonal to $\psi^P(p^\mu,\lambda)$ are given by 
\begin{eqnarray}
\label{eq:APST}
\psi^{AP}(p^\mu,\lambda) = \left( \begin{array}{c} \psi_+(p^\mu,\lambda) \\ -\psi_-(p^\mu,\lambda) \end{array} \right),
\end{eqnarray}
because $ \psi^\dagger_+(p^\mu,\lambda) \psi_-(p^\mu,\lambda)= \psi^\dagger_-(p^\mu,\lambda)\psi_+(p^\mu,\lambda)$ is satisfied from $\psi_\pm(k^\mu,\lambda)= \psi_\mp(k^\mu,\lambda)$. 

The parity operation on the base spin states $\psi^P(p^\mu,\lambda)$ and $\psi^{AP}(p^\mu,\lambda)$ are represented by the $\pm \gamma^0=\pm\left( \begin{array}{cc}  0 & \mathbb{I}_{2s+1} \\ \mathbb{I}_{2s+1} & 0 \end{array} \right)$ matrix because
\begin{subequations}
\begin{eqnarray}
\label{eq:OPTP} 
\mathcal{P}\left( \begin{array}{c} \psi_+(p^\mu,\lambda) \\ \psi_-(p^\mu,\lambda) \end{array} \right) &=& 
\left( \begin{array}{c} \psi_+(\tilde{p}^\mu,\lambda) \\ \psi_-(\tilde{p}^\mu,\lambda) \end{array} \right)  
=\gamma^0 
\left( \begin{array}{c} \psi_+(p^\mu,\lambda) \\ \psi_-(p^\mu,\lambda) \end{array} \right), \\
\label{eq:OPTAP} 
\mathcal{P} \left( \begin{array}{c} \psi_+(p^\mu,\lambda) \\ - \psi_-(p^\mu,\lambda) \end{array} \right)&=& 
\left( \begin{array}{c} \psi_+(\tilde{p}^\mu,\lambda) \\ -\psi_-(\tilde{p}^\mu,\lambda) \end{array} \right)   
= 
- \gamma^0 
\left( \begin{array}{c} \psi_+(p^\mu,\lambda) \\ -\psi_-(p^\mu,\lambda) \end{array} \right),
\end{eqnarray} 
\end{subequations}
where $\mathbb{I}_{2s+1}$ is the $(2s+1)$-dimensional identity matrix. 
The meaning of the superscripts $P$ and $AP$ will be clear in the context to come.

Let the spin operator in the parity-extended Poincar\'e group be ${S}^k=S^k_+\oplus S^k_-$, whose spin state representation becomes 
\begin{eqnarray}
\label{eq:DSSP}
S^k(p^\mu)=\left( \begin{array}{cc} S^k_+(p^\mu) & 0 \\ 0 & S^k_-(p^\mu) \end{array} \right) .
\end{eqnarray}
Under the parity operation the $S^k(p^\mu)$ transforms as
\begin{eqnarray}
S^k(p^\mu) \rightarrow S^k(\tilde{p}^\mu)=\gamma^0 S^k(p^\mu) \gamma^0,
\end{eqnarray}
whose eigenvalue is preserved under the parity operation as
\begin{eqnarray}
\label{eq:SPDSP} \nonumber
\mathcal{P} \left[{S}^k(p^\mu) \psi^{P/AP}({p}^\mu,\lambda)\right]&=& \pm \lambda \gamma^0  \psi^{P/AP}({p}^\mu,\lambda)=\lambda  \mathcal{P} \psi^{P/AP}({p}^\mu,\lambda)
\end{eqnarray}
using Eqs. (\ref{eq:OPTP}) and (\ref{eq:OPTAP}). 
Obviously, the Casimir operators are $P^\mu P_\mu$ and
 ${S}^k{S}_k$
 in the parity-extended Poincar\'e group. 

\subsection{Fundamental dynamical equation for free massive spin $1/2$ fields}
\label{subsec:FDE}

 In this subsection we will show that for spin $1/2$, the parity operation derives the fundamental dynamical equations for the spinor $\psi^{P/AP}(p^\mu, \lambda)$, which are equal to the usual free covariant Dirac equations for particle and antiparticle. It should be noted that the connection between the wave equation and the Poincar\'e group was discussed by Bargmann and Wigner \cite{Bargmann}, and the derivation of the Dirac equation based on the Bargmann-Wigner classification was given by Wightman \cite{Wightman}. More recently Thaller showed the derivation of the Dirac equation using the invariance of the positive and the negative energy spaces under the Lorentz boost \cite{Thaller}. The derivation by parity operation, however, is helpful to understand the identification of $\psi^{P}$ and $\psi^{AP}$ as particle and antiparticle states. 

The parity operation makes the direct sum $(s,0)\oplus(0,s)$ representation irreducible. This is because the parity operation acting on the spin state $\psi^{P/AP}(p^\mu, \lambda)$ can be represented by $\pm\gamma^0$ with off-diagonal component as we have seen in the previous section. The spin state $\psi^{P/AP}(p^\mu,\lambda)$ in the moving frame is obtained by using the standard LT as
\begin{eqnarray}
\label{eq:LTPAP}
\psi^{P/AP}(p^\mu, \lambda) = \mathnormal{Q}({\bf p})\psi^{P/AP}(k^\mu, \lambda)
\end{eqnarray}
from Eqs. (\ref{eq:PLT}) and (\ref{eq:APST}).
Then the parity operation on the $\psi^{P/AP}(p^\mu,\lambda)$ can be rewritten as 
\begin{eqnarray}
\mathcal{P}\psi^{P/AP}(p^\mu, \lambda)=\mathnormal{Q}^{-2}({\bf p}) \psi^{P/AP}(p^\mu, \lambda),
\end{eqnarray}
with 
\begin{eqnarray}
\mathnormal{Q}({\bf p})=e^{\gamma^5 {\bf S}(k^\mu)\cdot \mathbf{\zeta}}
\end{eqnarray}
using the spin operator in Eq. (\ref{eq:DSSP}) at the RF, 
where $\gamma^5 = \left( \begin{array}{cc} \mathbb{I} &0 \\ 0 & - \mathbb{I} \end{array} \right)$ and $ \mathbb{I}$ is the 2-dimensional identity matrix. 
This expression of the parity operation provide two different relations for $P$ and $AP$, respectively, as 
\begin{subequations}
\begin{eqnarray}
\label{eq:EQPPAT}
\gamma^0 \, \psi^P(p^\mu, \lambda)&=& \mathnormal{Q}^{-2}({\bf p})  \, \psi^P(p^\mu, \lambda) \\ 
\label{eq:EQAPPAT}
\gamma^0 \, \psi^{AP}(p^\mu, \lambda)&=& -\mathnormal{Q}^{-2}({\bf p})  \, \psi^{AP}(p^\mu, \lambda),
\end{eqnarray}
\end{subequations}
using Eqs. (\ref{eq:OPTP}) and (\ref{eq:OPTAP}). 

The exponential form of $\mathnormal{Q}^{-2}({\bf p}) $ becomes the linear form of $p^\mu$ as follows
\begin{equation}
  \mathnormal{Q}^{-2}({\bf p})   
 = \frac{1}{m}\!\left[ p^0 + 2\!\left( \begin{array}{cc} -{S}^{k}_+(k^\mu) p^k& 0 \\ 0
&  {S}^{k}_-(k^\mu) p^k \end{array} \right)\right],
  \end{equation}
	because the generators of the $SU(2)$ group satisfy $4 S^i S^j =\delta_{ij}+2i\epsilon_{ijk}S^k$ for spin $1/2$. 
Equations (\ref{eq:EQPPAT}) and (\ref{eq:EQAPPAT}) are then given as
\begin{subequations}
 \begin{eqnarray}
\label{eq:CPAO}
\gamma^0 \psi^P(p^\mu,\lambda)&=&\frac{1}{m}\!\left[ p^0 - 2\gamma^5\! {S}^{k}(k^\mu) p^k \right] \! \psi^P(p^\mu,\lambda) \\
\label{eq:CAPAO}
\gamma^0 \psi^{AP}(p^\mu,\lambda)&=&-\frac{1}{m}\!\left[ p^0 - 2\gamma^5\! {S}^{k}(k^\mu) p^k \right] \! \psi^{AP}(p^\mu,\lambda),
\end{eqnarray}
\end{subequations}
where $\gamma^0$ and $\gamma^5$ are the usual 4-dimensional gamma matrices in the chiral representation.
 By taking a parity operation again to Eqs. (\ref{eq:CPAO}) and (\ref{eq:CAPAO}) using $\pm\gamma^0$, 
we obtain the Lorentz covariant equations
\begin{subequations}
\begin{eqnarray}
\label{eq:FOESHP}
 \left( \gamma^0 p_0 + 2\gamma^0 \gamma^5 {S}^k(k^\mu) \; p_k - m \right) \psi^P(p^\mu,\lambda)=0,\\
\label{eq:FOESHAP}
\left( \gamma^0 p_0 + 2\gamma^0 \gamma^5 {S}^k(k^\mu) \; p_k + m \right) \psi^{AP}(p^\mu,\lambda)=0.
\end{eqnarray}
\end{subequations}
The $2\gamma^0 \gamma^5 {S}^k(k^\mu)$ is nothing but the Dirac gamma matrices 
\begin{eqnarray}
\gamma^k=\gamma^0\gamma^5 \Sigma^k,
\end{eqnarray}
where $\Sigma^k$ are 4-dimensional Pauli matrices. Then these equations are equal to the covariant Dirac equations for particle and antiparticle spinors \cite{Schwartz}. Note that the spin tensor operator $S^{\mu\nu}(p^\mu)$ determined by Eq. (\ref{eq:DSP2}) can be represented as 
	\begin{eqnarray}
	\label{eq:SPGAMMA}
S^{\mu\nu}(p^\mu)=\mathnormal{Q}({\bf p})\frac{i}{4}[{\gamma}^\mu, {\gamma}^\nu]\mathnormal{Q}^{-1}({\bf p}).
	\end{eqnarray}
	This relation becomes the usual one as 
	\begin{eqnarray}
	\label{eq:RFGAMMA}
	\Sigma^{\mu\nu}=\frac{i}{2}[{\gamma}^\mu, {\gamma}^\nu]
	\end{eqnarray}
	at the RF.

$\psi^P(p^\mu,\lambda)$ and $\psi^{AP}(p^\mu,\lambda)$ are the positive energy solutions of Eqs. (\ref{eq:FOESHP}) and (\ref{eq:FOESHAP}) and correspond to particle and antiparticle spinors, respectively. There are pair of negative energy solutions other than $\psi^{P/AP}(p^\mu,\lambda)$ in Eqs. (\ref{eq:FOESHP}) and (\ref{eq:FOESHAP}), however, these negative energy solutions are not proper states in the direct-sum $(1/2,0)\oplus(0,1/2)$ representation, because they cannot be obtained by the standard LT from the spin state at the RF. 
Equations (\ref{eq:FOESHP}) and (\ref{eq:FOESHAP}) can be represented as the following one equation
\begin{eqnarray}
\label{eq:CFDE}
\left( \gamma^0 P_0 + 2\gamma^k \; P_k - m \right) \Psi^{P/AP}(p^\mu,\lambda)=0
\end{eqnarray} 
by using 
\begin{subequations}
\begin{eqnarray}
\label{eq:PSTATE}
P^\mu \Psi^P(p^\mu,\lambda) &=& p^\mu \psi^P(p^\mu,\lambda)|p^\mu\rangle \\
\label{eq:APSTATE}
P^\mu \Psi^{AP}(p^\mu,\lambda) &=& -p^\mu \psi^{AP}(p^\mu,\lambda)|-p^\mu\rangle. 
\end{eqnarray}
\end{subequations}
Therefore, we obtained the fundamental dynamical equation for a free massive spin-$1/2$ field. 

Eq. (\ref{eq:CFDE}) manifests that the spin
 ${S}^k(k^\mu)$ is naturally included in 
the fundamental dynamical equation
 for free massive spin-$1/2$ fields. 
%
%
%
%
We have shown that the appearance of the Pauli matrices in the Dirac equation
 is a natural consequence of the helicity operator
 ${\bf S}(k^\mu) \cdot \hat{\bf p}$ written by the rest spin operator ${\bf S}(k^\mu)$. 
This fact could explain why the Dirac equation predicts the existence of spin
 and describes spin-$1/2$ massive elementary fields successfully.

%
%

\section{Spin as an observable}
\label{sec:Nthr}
 Of important issue is whether the spin ${\bf S}$ of free massive (non-chiral) spin-$1/2$ fields is a conserved quantity,
 because the Dirac spin $\boldsymbol{\Sigma}/2$ is not conserved. In general, Noether's method \cite{Noether} allows us to answer on this question explicitly.
 The new spin-$1/2$ Lagrangian, which gives the fundamental dynamical equation in Eqs. (\ref{eq:FOESHP}) and (\ref{eq:FOESHAP}), i.e., (\ref{eq:CFDE}) as the equation of motion, becomes the usual Dirac Lagrangian: 
\begin{eqnarray}
\label{eq:LAGOLD}
 {\cal L} = \bar\psi(x^\mu) (i\gamma^\mu \partial_\mu-m)\psi(x^\mu),
\end{eqnarray} 
with $P_\mu=i\partial_\mu$. 
Here the spinor field $\psi(x^\mu)$ is given from the Fourier representation as
\begin{eqnarray} 
\psi(x^\mu) 
=\sum_\lambda\int \frac{dp^3}{(2\pi)^3} \frac{m}{{p^0}} \left( a(p) e^{-i p^\mu x_\mu} \psi^P(p^\mu,\lambda)+b^*(p)e^{i p^\mu x_\mu} \psi^{AP}(p^\mu,\lambda)\right),
\end{eqnarray}
where $a(p)$ and $b^*(p)$ are arbitrary Lorentz-invariant complex scalar function of $p^\mu$. 
We will denote $\psi(x^\mu)$ as $\psi$ for simplicity. 
Even though ${\cal L} $ is equal, the differences of the representation of the Lorentz transformation given by the two different spin operators, i.e., the spin ${\bf S}(p^\mu)$ and the Dirac spin $\boldsymbol{\Sigma}/2$, will provide different Noether conserved currents under the Lorentz symmetry that determine conserved quantities. 

\subsection{Conserved Noether currents and charges}

 Let us first review the conventional Dirac case in which the Lorentz transformation is represented by
\begin{eqnarray}
\exp{(i\omega_{\mu\nu}\Sigma^{\mu\nu}/2)},
\end{eqnarray}
where $\epsilon_{ijk}\omega^{ij}/2$ and $\omega^{i0}$ correspond to the rotation angle $\theta^k$ and the rapidity $\xi^i$, respectively, and the spin matrices are $\Sigma^{\rho\sigma} = \frac{i}{4}\left[ \gamma^\rho,
 \gamma^\sigma\right]$ that gives $\Sigma^k=\frac{1}{2}\epsilon_{ijk}\Sigma^{jk}$.
 For the Lorentz invariance,
 Noether's theorem \cite{Noether} gives
 the conserved currents as
\begin{equation}
 ({\cal J}_{\scriptsize{D}}^\mu)^{\rho\sigma}
 = x^\rho T^{\mu\sigma} - x^\sigma T^{\mu\rho}
  +\bar\psi \gamma^\mu \frac{\Sigma^{\rho\sigma}}{2}\psi,
  \label{eq:Totalcurrent}
\end{equation}
where the energy-momentum tensors of the spinor field $\psi$ are
 $T^{\mu\nu} = i\bar\psi \gamma^\mu \partial^\nu \psi$.
 The conserved currents give rise to the conserved charges corresponding to the total angular momentum \cite{tong}
\begin{eqnarray}
 q^{ij} = \int d^3x ({\cal J}_{\scriptsize{D}}^0)^{ij}= \int d^3x \, \left( x^i T^{0j} - x^j T^{0 i}
  +\bar\psi \gamma^0 \frac{\Sigma^{ij}}{2}\psi \right)
	\end{eqnarray}
  and the conserved quantities under pure boosts
	\begin{eqnarray}
  q^{0i} = \int d^3x ({\cal J}_{\scriptsize{D}}^0)^{0i}= \int d^3x \, \left( x^0 T^{0i} - x^i T^{0 0}
  +\bar\psi \gamma^0 \frac{\Sigma^{0i}}{2} \psi \right).
	\end{eqnarray}
 The conserved Noether currents satisfy $\partial_\mu ({\cal J}^\mu)^{\rho\sigma} =0$. However, one can confirm that the last (spin) term of Eq.
 (\ref{eq:Totalcurrent}) 
 does not satisfy itself
 the current conservation, i.e., 
\begin{eqnarray}
\partial_\mu (\bar\psi \gamma^\mu
 \frac{\Sigma^{\rho\sigma}}{2}\psi) \neq 0.
\end{eqnarray}
This fact implies that the current associated with the Dirac spin is not conserved by itself and then the Dirac spin is not conserved quantity. This result is consistent with the fact that the Dirac spin does not commute with the Dirac Hamiltonian \cite{Dirac}.

 Next let us consider the Lorentz transformation represented by the spin $S^{\mu\nu}$ as
\begin{eqnarray}
\exp{(i\omega_{\mu\nu}S^{\mu\nu})}
\end{eqnarray} 
with $S^{\mu\nu}\Psi^{P/AP}(p^\mu, \lambda)=S^{\mu\nu}(p^\mu)\Psi^{P/AP}(p^\mu, \lambda)$ for $S^{\mu\nu}(p^\mu)$ in Eq. (\ref{eq:SPGAMMA}). 
The corresponding Noether conserved currents become
\begin{equation}
 ({\cal J}^\mu)^{\rho\sigma}
 = { X}^\rho T^{\mu\sigma} - { X}^\sigma T^{\mu\rho}
  + \bar\psi {\gamma}^\mu { S}^{\rho\sigma}\psi
 \label{eq:Totalcurrent2}
\end{equation}
where ${ X}^\mu$ is the new position operator corresponding to the spin $S^{\mu\nu}$. The new position operator preserves the commutation relation with the momentum. Note that in the representation $P^\mu=i\frac{\partial}{\partial x^\mu}$ with the usual position operator $x^\mu$, there is an ambiguity to determine a position operator satisfying $[X^\mu, P^\nu]=-ig^{\mu\nu}$, which admits $X^i= x^i +f^i(P^\mu)$, where $f^i(P^\mu)$ is the function of $P^\mu$ and transforms as the $i$-component of a 4-vector under Lorentz transformation. This ambiguity $f^i(P^\mu)$ can be fixed by using the equality of Eqs. (\ref{eq:Totalcurrent}) and (\ref{eq:Totalcurrent2}), which requires that the total angular momentum should be equal independent on the different decompositions into orbital and spin angular momentum, and the locality condition $[{ X}^\mu, { X}^\nu] =0$ in Refs. \cite{NW,Fleming64} with ${X}^0=x^0$. As a result, the new position operator satisfies $[{ X}^\mu, {S}^k]=0$ similar to $[x^\mu, \Sigma^k/2]=0$. 

Let us define the spin-associated current from the total conserved current in Eq. (\ref{eq:Totalcurrent2}) as
\begin{equation}
 \label{eq:Scurrent}
 \left( {\cal J}^{\mu}_S\right)^{\rho\sigma}
  = \bar\psi {\gamma}^\mu { S}^{\rho\sigma} \psi. 
\end{equation}
 To be conserved itself, the spin-associated current $ \left( {\cal J}^{\mu}_S\right)^{\rho\sigma}$ should satisfy the conservation condition, i.e., 
\begin{eqnarray}
\label{eq:SCCON}
\partial_\mu \left( {\cal J}^{\mu}_S\right)^{\rho\sigma} =
\left( \partial_\mu \bar\psi \right) {\gamma}^\mu { S}^{\rho\sigma} \psi 
+ \bar\psi {\gamma}^\mu \partial_\mu \left( { S}^{\rho\sigma} \psi \right)=0.
\end{eqnarray}
Using the fundamental dynamical equations (\ref{eq:FOESHP}) and (\ref{eq:FOESHAP}), the conservation condition (\ref{eq:SCCON}) is shown to be equivalent to the following commutation condition,
\begin{eqnarray}
 [ {\gamma}^{\alpha} p_{\alpha}, {S}^{\rho\sigma}(p^\mu)] \psi^{P/AP}(p^\mu,\lambda)=0.
\end{eqnarray} 
Then one can justify $\partial_\nu \left( {\cal J}^{\nu}_S\right)^{ij} =0$ as follows,
\begin{eqnarray}
[ {\gamma}^{\nu} p_{\nu}, {S}^{ij}(p^\mu)]\psi^{P/AP}(p^\mu,\lambda) &=&
[ {\gamma}^{\nu} p_{\nu}, {S}^{ij}(p^\mu)] \psi^{P/AP}(p^\mu,\lambda) \\ \nonumber
&=& \frac{1}{2}{\mathnormal{Q}}({\bf p}) [ \gamma^\delta k_\delta, \Sigma^{ij}] {\mathnormal{Q}}^{-1}({\bf p})\psi^{P/AP}(p^\mu,\lambda) \\ \nonumber
&=& 0
\end{eqnarray}
with Eqs. (\ref{eq:LTPAP}) and (\ref{eq:SPGAMMA}) and the transformation of the gamma matrices, i.e.,
\begin{eqnarray}
{\mathnormal{Q}}^{-1}({\bf p}) \gamma^\mu {\mathnormal{Q}}({\bf p}) = L^\mu_{\phantom{\mu}\delta}\gamma^\delta=
 (L^{-1})_\delta^{\phantom{\mu}\mu}\gamma^\delta ,
\end{eqnarray}
with $p^\mu=L^\mu_{\phantom{\mu}\nu}k^\nu$. 
%
This implies that the spin current $\left( {\cal J}^{\mu}_S\right)^{ij}$ is conserved by itself. 
However, one can find $\partial_\nu\left( {\cal J}^{\nu}_S\right)^{0i}\neq 0$ from
 \begin{eqnarray}
[  {\gamma}^{\nu} p_{\nu} , {S}^{0i}(p^\mu) ]\psi^{P/AP}(p^\mu,\lambda) \neq 0
 \mbox{ because } [  m \gamma^0, \Sigma^{0i}] \neq 0.
\end{eqnarray}
 This shows that under pure boosts,
 only the total current $({\cal J}^\mu)^{0i}$ is conserved itself.

 Based on the conserved spin current $\left( {\cal J}^{\mu}_S\right)^{ij}$, the total current $({\cal J}^\mu)^{ij}$, giving rise to the total angular momentum
 as the conserved charge,
 can be decomposed into the two conserved
 currents as
\begin{subequations}
 \begin{equation}
 ({\cal J}^\mu)^{ij} = ({\cal J}^\mu_L)^{ij} + ({\cal
 J}^\mu_S)^{ij},
 \end{equation}
 where the orbital current
 $({\cal J}^\mu_L)^{ij}$ and the spin current $({\cal J}^\mu_S)^{ij}$
 given by
 \begin{eqnarray}
 ({\cal J}^\mu_L)^{ij} &=& { X}^i T^{\mu\, j} - { X}^j T^{\mu\, i},
 \label{eq:ScurrentL}
 \\
 ({\cal J}^\mu_S)^{ij} &=& \bar\psi {\gamma}^\mu { S}^{ij}\psi.
 \label{eq:ScurrentS}
 \end{eqnarray}
\end{subequations}
Obviously, the orbital current satisfies its conservation condition, i.e., $\partial_\mu({\cal J}^\mu_L)^{ij}=0$. The orbital and the spin current
 give rise to the corresponding conserved charges, i.e.,
 the orbital angular momentum of the spinor field $\psi$
\begin{eqnarray}
 \mathcal{L}^{ij} = \int d^3x ({\cal J}^0_L)^{ij}= \int d^3x \left( { X}^i T^{0\, j} - { X}^j T^{0\, i} \right)
\end{eqnarray}
 and the spin angular momentum of the spinor field $\psi$
\begin{eqnarray}
 \mathcal{S}^{ij} = \int d^3x ({\cal J}^0_S)^{ij} = \int d^3x\, \psi^\dagger { S}^{ij}
 \psi,
\end{eqnarray}
respectively. 
Using the $\mathnormal{S}^{ij}$, the conserved spin three-vector of the spinor field $\psi$ is given as
\begin{eqnarray}
\label{eq:CSTV}
\mathcal{S}^k= \frac{1}{2}\epsilon_{ijk} \mathcal{S}^{ij} =  \int d^3x\, \psi^\dagger { S}^{k} \psi.
\end{eqnarray}
%
%

\subsection{Good observable spins for free particles and antiparticles} 

Of particular importance is whether spin is an observable or not. To be an observable, spin should be Hermitian. 
One can notice that the $\mathcal{S}^k$ in Eq. (\ref{eq:CSTV}) cannot be observed  because the spin operator ${\mathcal S}^{k}$ is not Hermitian. However, when we measure spin, we observe it for either a particle or an antiparticle, not a field itself. This implies that as a good observable, spin could be defined for a particle or an antiparticle separately. In our representation space, the particle and the antiparticle projection operators $\Pi^{P/AP}$ can be introduced as 
\begin{eqnarray}
\Pi^{P}= \frac{m + {\gamma}^\mu p_\mu }{2m} \mbox{ and } \Pi^{AP}= \frac{m - {\gamma}^\mu p_\mu }{2m}
\end{eqnarray}
using the fundamental dynamical equations (\ref{eq:FOESHP}) and (\ref{eq:FOESHAP}).
 Then the particle and the antiparticle spinors $\psi^P$ and $\psi^{AP}$ are given by $\psi^{P/AP}=\Pi^{P/AP}\psi$. 
Consequently, one can define the conserved spin (angular momentum) for a particle and an antiparticle as
\begin{subequations}
\begin{eqnarray}
\label{eq:PSP}
\mathbf{\mathcal{S}}^{P}=\int d^3 x ~ \psi^{P\dagger} {\bf S}   \psi^P  \\
\label{eq:APSP}
\mathbf{\mathcal{S}}^{AP}=\int d^3 x ~ \psi^{AP\dagger} {\bf S}   \psi^{AP},
\end{eqnarray} 
\end{subequations}
respectively.

It will be shown that the spin $\mathbf{\mathcal{S}}^{P/AP}$ for particles and antiparticles are Hermitian in the following:
 The spin operators ${\bf S}(p^\mu)$ in the spin state representation are expressed by the following similarity transformation from Eq. (\ref{eq:SPGAMMA}) as
\begin{eqnarray}
{\bf S}(p^\mu)= {\mathnormal{Q}}({\bf p}) \frac{\boldsymbol{\Sigma}}{2}{\mathnormal{Q}}^{-1}({\bf p})
\end{eqnarray}
and the projection operator satisfies
\begin{eqnarray}
\label{eq:PNUT}
\Pi^{P/AP}\psi(p^\mu,\lambda)= {\mathnormal{Q}}({\bf p})\frac{1\pm \gamma^0}{2}\psi(k^\mu,\lambda) 
= {\mathnormal{Q}}^{P/AP}({\bf p}) \frac{1\pm \gamma^0}{2}\psi(k^\mu,\lambda)
\end{eqnarray}
with the defined matrices 
\begin{eqnarray}
\label{eq:UTPAP}
{\mathnormal{Q}}^{P/AP}({\bf p})=e^{\pm \gamma^5\gamma^0 \boldsymbol{\Sigma}\cdot{\mathbf{\zeta}}/2},
\end{eqnarray}
where $+$ and $-$ in Eqs. (\ref{eq:PNUT}) and (\ref{eq:UTPAP}) correspond to the superscript $P$ and$AP$, i.e., particle and antiparticle, respectively. The unitary matrix ${\mathnormal{Q}}^{P/AP}({\bf p})$ can be re-expressed by using the unitary matrices $U^{P/AP}({\bf p})$ as
\begin{eqnarray}
{\mathnormal{Q}}^{P/AP}({\bf p})= U^{P/AP}({\bf p}),
\end{eqnarray} 
where $\left(U^P({\bf p})\right)^\dagger$ is the Foldy-Woutheysen (FW) transformation matrix \cite{FW} and $U^{AP}({\bf p})=\left(U^{P}({\bf p})\right)^{-1}$. 
Accordingly, the following relation holds
\begin{eqnarray}
\label{eq:GURS}
{\bf S}(p^\mu)\psi^{P/AP}(p^\mu,\lambda) &=&U^{P/AP}({p^\mu})\frac{\boldsymbol{\Sigma}}{2} \left(U^{P/AP}\right)^\dagger({p^\mu})\psi^{P/AP}(p^\mu,\lambda) \\ \nonumber
&=& {\bf S}^{P/AP}(p^\mu) \psi^{P/AP}(p^\mu,\lambda)
\end{eqnarray}
by defining the particle and the antiparticle spin operators 
\begin{eqnarray}
{\bf S}^{P/AP}(p^\mu) :=U^{P/AP}({\bf p})\frac{\boldsymbol{\Sigma}}{2} \left(U^{P/AP}({\bf p})\right)^\dagger({p^\mu}).
\end{eqnarray}  
Then the ${\bf {S}}^{P/AP}(p^\mu)  $ are definitely Hermitian and as a result, the $\mathbf{\mathcal{S}}^{AP}$ become Hermitian and good observables. Interestingly, one may notice that the helicity operator satisfies 
\begin{eqnarray}
{\bf S}(p^\mu)\cdot \hat{\bf p}=\frac{1}{2}\boldsymbol{\Sigma}\cdot \hat{\bf p},
\end{eqnarray}
where $\hat{\bf p}={\bf p}/|{\bf p}|$ is the unit vector in the direction of the momentum. This shows that the helicity can be a good observable even though the Dirac spin itself is not. 

Usually, spin is considered as an axial vector \cite{Bogolubov} because the classical angular momentum is an axial vector. In this sense, as a correspondence, the axial vector condition can be considered as a consistency condition for spin. The spin ${\bf S}$, due to the fact that the parity operation does not change the spin eigenvalue, can be considered as an axial vector. 
This fact is also confirmed in the conserved spin angular momentum. Under parity, the spin states transform as $\psi(x^0,{\bf x}) \rightarrow \pm\gamma^0 \psi(x^0, -{\bf x})$ according to whether $\psi(x)$ is a particle or an antiparticle. The invariance of spin under parity can be shown as 
\begin{eqnarray}
\int d^3x\,\psi^\dagger {S}^k \psi \rightarrow \int d^3 x \psi^\dagger\gamma^0 \gamma^0 {S}^k  \gamma^0 \gamma^0 \psi
\end{eqnarray} 
with $(\gamma^0)^\dagger =\gamma^0$. As a result, the conserved spin angular momentum ${\mathbf{\mathcal S}}  $ also becomes axial. 
Therefore, as a physical observable, the spin for a particle and an antiparticle is Hermitian and axial as usually expected. One can also confirm easily that the parity-inversion spin state
 $\pm\gamma^0 \psi(x^0,-{\bf x})$ satisfy the same fundamental equations (\ref{eq:FOESHP}) and (\ref{eq:FOESHAP}). Note that the relation for the particle state in Eq. (\ref{eq:GURS}) is equivalent to that in Ref. \cite{Gursey}.


\section{Conclusion and Discussion}
\label{sec:Conclusion}

%
%
 We have derived the two spin operators, whose squares become the second Casimir invariant of the Poincar\'e group, from a general linear combination of the components of the PL vector by imposing the minimum requirements. 
The two spin operators admit the complexified $\mathfrak{su}(2)$ algebra, which provide the two inequivalent, i.e., the left- and the right-handed spin state representations of the Lorentz generators. The two spin operators do not commute with each other, hence the representations are not equal to those for the homogeneous Lorentz group, which are the tensor product of two spin angular momentum.   
%
 %
As a result, the left-handed and the right-handed representation of the Poincar\'e group has no trivial scalar representation that has no clear physical meaning, although the mathematical expressions for the representation space are the same both for the homogeneous and inhomogeneous Lorentz group. The left-handed and the right-handed representations are covariant different from that of the unitary irreducible representation.
 However, the two spin operators are not Hermitian, which indicates that a physical observable theory cannot be provided by themselves. 

 Under the parity transformation, the two spin operators exchange each other, which leads to the fact that the parity operation transforms the left-handed representation to the right-handed representation and vice versa. 
In consequence, the only natural irreducible representation without any redundant representation space under the parity-extended Poincar\'e group is the direct sum $(s,0)\oplus(0,s)$ representation corresponding to free massive elementary non-chiral spin $s$ fields. In this direct sum representation we constructed two kinds of base states that are orthogonal to each other under the Lorentz invariant scalar product and have parity eigenvalue $\pm 1$ at the RF. These two kinds of states are shown to satisfy the fundamental dynamical equations for $(1/2,0)\oplus(0,1/2)$ representation, which are identical to the covariant Dirac equations for 
free massive elementary non-chiral spin $1/2$ particles and antiparticles. The aprity operation plays a crucial role to derive the covariant Dirac equations, which implies that the spacetime symmetry for the free Dirac field is parity-extended Poincar\'e group.  
%
%
%

In contrast to previous approaches suggesting relativistic spin operators for spin-$1/2$ massive particles,
 we 
 have enabled to manifest from the Noether's theorem that the spin angular momentum $\int d^3x\, {\psi}^\dagger{\bf S}\psi$ is a conserved quantity and the Dirac spin angular momentum $\int d^3x\, {\psi}^\dagger({\boldsymbol{\Sigma}}/2) \psi$ is not. 
The conserved spin angular momentum $\int d^3x\,{\psi}^\dagger {\bf S}\psi$ is also an axial vector as usually expected for a spin angular momentum but it is still not Hermitian, however, its projected form $\int d^3 x \psi^{P/AP\dagger}{\bf S}\psi^{P/AP} $ either on the particle or on the anti-particle space has been shown to be Hermitian. This fact suggests that one can measure either a particle or an antiparticle only, not a field itself in physical measurements. 
%
%
 %

The definition of spin can be used to define the position operator through the total angular momentum. That is, the canonical position operator and the new position operator correspond to the Dirac spin and the new spin, respectively. We recently studied the different motions expected from the Dirac and the new position operators, which can be verified in high energy electron beam experiment \cite{Ours1}. This experiment will also show which spin is good observable.

%


\acknowledgments

We acknowledge support from the National Research Foundation of Korea Grant funded by the Korea Government (2016-0113, 2016-0234, T.C.)
 and the National Natural Science Foundation of China under the Grant
 No. 11374379 (S. Y. C.). T.C. thanks Y.D. Han for helpful discussions.

\section{appendix}
\subsection{ Derivation of spin operator with minimum requirements}

We will show the details of the derivation for the spin operators using the following three requirements:

$\bullet$ $S^k$ should commute with the momentum; $[S^k, P^\mu]=0$ (little group condition).

 $\bullet$ $S^k$ should satisfy the $\mathfrak{su}(2)$ algebra; $[S^i, S^j]=i\epsilon_{ijk} S^k$ (angular momentum condition).

 $\bullet$  $S^k$ should transform as a $k0$-component of a second-rank tensor under Lorentz transformation (tensor condition).

(i) To show the differences from the Bogolubov {\it et al.}'s derivation clearly, we will start with Bogolubov {\it et al.}'s vector condition iv):
 \begin{equation}
  [ J^j, S^k ] = i\epsilon_{jkl} S^l ,
  \label{eq:rotation}
 \end{equation}
 where $J^j=\epsilon_{jlm}J^{lm}/2$ is the rotation generator around the axis $\hat x^j$. In fact, the vector condition is the three dimensional notation of the requirement that $S^k$ should transform as a $k0$-component of the second-rank dual spin tensor under rotation, i.e,
\begin{eqnarray}
\frac{1}{2}\epsilon_{jlm} [ J^{lm}, {\ast S}^{k0}]=i \epsilon_{jkl} {\ast S}^{l0}, 
\end{eqnarray}
for
 $J^j = \epsilon_{jlm} J^{lm}/2$.
The tensor condition additionally restricts the transformation of $S^k$ under the boost transformation.  
 Eq. (\ref{eq:rotation}) with $S^k$ in Eq. (\ref{eq:linear}) gives the relation 
  \begin{equation}
  [ J^j, a_{k,\mu} ] W^\mu + i\epsilon_{jln} a_{k,l}  W^n =  i\epsilon_{jkl} a_{l,\nu} W^\nu
  \label{eq:equal1}
 \end{equation}
 by using
 $[ J^{\lambda\mu}, W^\nu ] = i (g^{\mu\nu} W^{\lambda} - g^{\lambda\nu} W^{\mu})$.
 Since all $W^\mu$ terms can be considered to be linearly independent (the constraint $W^\mu P_\mu=0$ does not apply to determine the coefficient $a_{k,\mu}$ because $W^\mu P_\mu=0$ is automatically satisfied \cite{Note1}),
 the coefficients $a_{k,\mu}$ in Eq. (\ref{eq:equal1}) should satisfy

\begin{subequations}
 \begin{eqnarray}
 \left[ J^j, a_{k,0} \right] = i \epsilon_{jkl} a_{l,0}
 \mbox{~\hspace{0.1cm} for~} W^0,
 \label{eq:condition0}
 \\
 \left[ J^j, a_{k,k} \right]  + i \epsilon_{jlk} a_{k,l} = i \epsilon_{jkl} a_{l,k}
 \mbox{~\hspace{0.1cm} for~} W^k,
 \label{eq:conditionK}
 \\
 \left[ J^j, a_{k,m \neq k} \right] + i \epsilon_{jl(m\neq k)} a_{k,l}
    = i \epsilon_{jkl} a_{l,m\neq k}
 \mbox{~for~} W^m.
 \label{eq:conditionM}
 \end{eqnarray}
\end{subequations}

 As a function of the momentum operator given by the little group condition,
 the coefficient $a_{k,0}$ in Eq. (\ref{eq:condition0})
 should be a function of  $P^k$ and $P^0$ because for $j=k$, $[J^k, a_{k,0}]=0$ is guaranteed from $[J^k, P^0]=0$ and $[J^k,P^k]=0$
 in the commutation relation $[J^{\mu\nu}, P^\rho] = i(g^{\nu\rho} P^\mu - g^{\mu\rho} P^\nu)$.
 In order to satisfy Eq. (\ref{eq:condition0}) for $j \neq k$, also,
 $a_{k,0}$ should be a linear function of $P^k$ because if it is quadratic
 or higher order functions of $P^k$ then
 the left-hand side of Eq. (\ref{eq:condition0}) becomes zero,
 but the right-hand side of Eq. (\ref{eq:condition0}) cannot be zero with general momentum.
 Then, the coefficient $a_{k,0}$ of the term $W^0$ should be written as
 \begin{equation}
  a_{k,0} = f_0(P^0) \ P^k,
 \label{eq:ak0}
 \end{equation}
  where $f_0(P^0)$ is a function of $P^0$.


 Eq. (\ref{eq:conditionM}) becomes
 $[ J^k, a_{k,m\neq k} ] = i\epsilon_{kml} a_{k,l}$ for $j =k$
 and $[ J^m, a_{k,m\neq k} ] = i\epsilon_{mkl} a_{l,m}$ for $j=m$.
 This implies that
 the non-commuting part of the operator $a_{k,m\neq k}$ with $J^l$
 transforms as the $m$- or $k$-component of a three-vector under a rotation.
 As for three-dimensional vector, only two types are possible.
 One is an ordinary vector $\mathbf{P}$,
 the other is a pseudovector $\mathbf{P}\times\mathbf{C}$
 with a constant vector $\mathbf{C}$.
 To satisfy Eq. (\ref{eq:conditionM}), then, the $a_{k,m\neq k}$ should be expressed as
 \begin{equation}
  a_{k,m\neq k} = f_2(P^0) \ P^k P^m + f_3(P^0) \epsilon_{kml} P^l,
 \label{eq:akm}
 \end{equation}
  where $f_2(P^0)$ and $f_3(P^0)$ are functions of $P^0$.


 The coefficient $a_{k,k}$ in Eq. (\ref{eq:conditionK})
 should be a function of $P^k$ and $P^0$ because $a_{k,k}$ commutes with $J^k$ for $j=k$.
 For $j \neq k$, furthermore,
 Eq. (\ref{eq:conditionK}) becomes $[ J^j, a_{k,k} ] =0$ by using the coefficient
 $a_{k,m\neq k}$ in Eq. (\ref{eq:akm}).
 Then, the coefficient $a_{k,k}$ should not be a linear function of $P^k$.
 And, for $j \neq k \neq m$, Eq. (\ref{eq:conditionM})
 can be $[J^j, a_{k,m\neq k} ] + i\epsilon_{jkm} a_{k,k}  = i\epsilon_{jkm} a_{m,m}$.
 Satisfying this condition,
 $a_{k,k}$ can have $f_1(P^0)$ or $f_2(P^0) P^k P^k$, so
\begin{equation}
  a_{k,k} = f_1(P^0) + f_2(P^0) P^k P^k,
 \label{eq:akk}
 \end{equation}
  where $f_1(P^0)$ is a function of $P^0$.

 Consequently, as a three-dimensional vector satisfying Eq. (\ref{eq:rotation}),
 $S^k$ in Eq. (\ref{eq:linear})
 can be rewritten in terms of a more specific form of the coefficients $a_{k,\mu}$ as
 \begin{eqnarray}
\label{eq:ASPIN} \nonumber
 S^k &=& f_0(P^0) P^k W^0 + f_1(P^0) W^k + f_2(P^0) P^k P^n W^n + \ f_3(P^0)\ \epsilon_{kml}\ P^l W^m,
 \end{eqnarray}
where we used 
\begin{eqnarray}\nonumber
f_2(P^0) P^k P^k W^k + \sum_{m\neq k} f_2(P^0) P^k P^m W^m &=& \sum_n f_2(P^0) P^k P^n W^n  \equiv  f_2(P^0) P^k P^n W^n.
\end{eqnarray}
The third term in Eq. (\ref{eq:ASPIN}) can be rewritten as 
\begin{eqnarray}
f_2(P^0) P^k P^n W^n=f_2(P^0) P^k P^0 W^0 =\tilde{f}_2(P^0)P^k W^0
\end{eqnarray}
by using $W^\mu P_\mu=0$ and $\tilde{f}_2=f_2 P^0$. Hence we can rewrite the $S^k$ in simpler form as 
\begin{eqnarray}
 S^k = f_4(P^0) P^k W^0 + f_1(P^0) W^k 
        + \ f_3(P^0)\ \epsilon_{kml}\ P^l W^m, 
 \label{sol1}
 \end{eqnarray}
where $f_4(P^0)= f_0(P^0)+\tilde{f}_2(P^0)$.


 (ii) The spin three-vector operators are
 \textit{generators of $\mathit{SU}(2)$ group}
  such that
  they should satisfy the $\mathfrak{su}(2)$ algebra, i.e., the commutation relations,
  \begin{equation}
  [ S^i, S^j ] = i\epsilon_{ijk} S^k.
  \label{eq:su2}
  \end{equation}
  Let us put $S^k$ in Eq. (\ref{sol1}) into the commutation relation in Eq. (\ref{eq:su2}).
  By using the commutation relations $[W^0, W^k] = i\epsilon_{klm} W^l P^m $
  and $[W^i,W^m] = i\epsilon_{iml} (W^l P^0  - W^0 P^l)$,
  the following three equations are obtained 
\begin{subequations}
 \begin{eqnarray}
 f_4 &=& - f_4\ f_1 P^0 - f_1^2 +  m^2 f_3^2,
 \label{eq:f1}
 \\
 f_1 &=& f_4 \ f_1 \ (P_0^2 - m^2) + f_1^2 \ P^0,
 \label{eq:f2}
 \\
 f_3 &=& f_4 \ f_3 \ (P_0^2 - m^2) + f_1 \ f_3 \ P^0.
 \label{eq:f3}
 \end{eqnarray}
\end{subequations}
 From Eqs. (\ref{eq:f1}), (\ref{eq:f2}), and (\ref{eq:f3}), however,
 $f$s, all of which are nonzero, cannot be determined because  Eqs. (\ref{eq:f2}) and (\ref{eq:f3}) are not independent each other for $f_1 \neq 0$ and $f_3 \neq 0$,
 which means that infinitely many solutions are possible with respect to $f$s.


 (iii) To specify $f$s more, we consider the fact that the spin angular momentum three-vectors
 are obtained from the second-rank tensors, i.e., $S^k=\ast S^{k0}$.
%
%
 Hence, $S^k$ should be transformed as
 a $k0$-component of second-rank tensor for a LT (tensor condition).

 Then, every term of $S^k$ in Eq. (\ref{sol1}) should satisfy the tensor condition separately. 
As a result, 
 $f_1$ should be linearly proportional to $P^0$, i.e., $f_1(P^0) = b\, P^0$,
 to make the term of $f_1(P^0)W^k$ transform like a $k0$-component of the tensor under an LT,
 while $f_4$ and $f_3$ should be constant (Lorentz scalar) because the terms of $P^k W^0$ and
 $\epsilon_{kml}P^l W^m = \epsilon_{0kml} P^l W^m$ already transform like
 a $k0$-component. Let $f_4(P^0) = a$
 and $f_3(P^0) = c$, then, $a$, $b$, and $c$ are Lorentz scalar. 
%

 Consequently,
 Eq. (\ref{sol1}) can be rewritten as a more specific form:
 \begin{equation}
 S^k = a\ P^k W^0 + b\ P^0 W^k + c\ \epsilon_{kml} P^l W^m.
 \label{newS}
 \end{equation}
 Then,
 Eqs. (\ref{eq:f1}), (\ref{eq:f2}), and (\ref{eq:f3}) become,
 respectively,
\begin{subequations}
 \begin{eqnarray}
 a  &=& - a\ b\ P_0^2 - b^2 P_0^2 + m^2\ c^2,
 \label{eq:f1a}
 \\
 b &=& a\  b\ (P_0^2 - m^2) + b^2\ P_0^2,
 \label{eq:f2b}
 \\
 c &=& a\  c\ (P_0^2 - m^2) + b\ c\ P_0^2.
 \label{eq:f3c}
 \end{eqnarray}
\end{subequations}
 Under boost transformation, $P^0$ transforms to different value, but $a$, $b$, and $c$ are invariant, hence both the coefficients of $P_0^2$
 and the constant terms in the three equalities $(\ref{eq:f1a})$, $(\ref{eq:f2b})$, and $(\ref{eq:f3c})$ should be zero separately.
 To determine the three constants $a$, $b$, and $c$, then,
 we obtain the six conditions:
\begin{subequations}
 \begin{eqnarray}
 && a\ (a + b) = 0 \mbox{~~~and~~~~~} a\ - m^2 \ c^2 =0,
 \\
 && b\ (a + b) = 0 \mbox{~~~and~~} b\ ( 1 + m^2\ a) = 0,
 \\
 && c\ (a + b) = 0 \mbox{~~~and~~~}  c\ ( 1 + m^2\ a) = 0.
 \end{eqnarray}
\end{subequations}
 These six conditions clearly show that if one of the three constants $a$, $b$, and $c$ is zero then
 all of the three constants become zero.
 Hence, all of them should be nonzero and then
 the six conditions reduce to
 the three conditions:
\begin{subequations}
 \begin{eqnarray}
  a + b &=& 0,
  \label{eq:three1}
 \\
  1 + m^2\ a  &=& 0,
 \label{eq:three2}
 \\
  a - m^2\ c^2 &=& 0.
  \label{eq:three3}
 \end{eqnarray}
\end{subequations}
 One can obtain the two sets of the three constants as
 \begin{equation}
 a =-\frac{1}{m^2}, \mbox{~~~} b = \frac{1}{m^2}, \mbox{~~and~~} c = \pm \frac{i}{m^2}.
 \end{equation}
 Resultantly, we obtain the two spin three-vectors in Eq. (\ref{eq:pm}) as
 \begin{equation}
 S^k_{\pm} = \frac{1}{m^2} \left( P^0 W^k - P^k W^0\right) \pm \frac{i}{m^2} \epsilon_{kml} P^l W^m. \phantom{------} 
 \nonumber
 \end{equation}

(iv) As a reference, we consider the Bogolubov {\it et al.}'s derivation in our derivation. 
They imposed the additional condition that the spin should be axial, which requires $f_3=0$ in Eqs. (\ref{eq:f1}), (\ref{eq:f2}), and (\ref{eq:f3}), instead of the tensor condition. 
Then there are two solution 
\begin{eqnarray}
\label{eq:BGAXIAL}
f_4=- \frac{1}{m(m\pm P^0)}, \mbox{~~~} f_1 =\pm\frac{1}{m}.
\end{eqnarray}
One of the two, which corresponds to the upper sign in Eq. (\ref{eq:BGAXIAL}), satisfies the condition that the three components of the spin operator reduces to the three spatial components of the PL vector at the RF 
such that the spin under the axial vector condition becomes
\begin{eqnarray}
\label{eq:BGL}
S^k_{\scriptsize{\mbox{\scriptsize{Bogolubov}}}}=\frac{W^k}{m}-\frac{W^0 P^k}{m(m+P^0)},
\end{eqnarray}
which is the same as the spin derived by Bogolubov {\it et al.} 
As one can easily check, with these $f_1$ and $f_4$ in Eq. (\ref{eq:BGAXIAL}) $S^k_{\scriptsize{\mbox{\scriptsize{Bogolubov}}}}$ cannot satisfy the tensor condition. Hence, $S^k_{\scriptsize{\mbox{\scriptsize{Bogolubov}}}}$ cannot be the component of a relativistically covaraint spin and in fact, it is the $k$-component of the PL vector transformed to the RF from the PL vector at the moving frame as Bogolubov {\it et al.} noted \cite{Bogolubov}. 
%
%

%
 \subsection{ Expression of spin using similarity transformation}
%
 The spin operator $S^k_+ ({p}^\mu)$ in Eq. (\ref{eq:spin})
 can also be expressed by the similarity transformation of the rest spin operator $\sigma^k /2$ as 
\begin{eqnarray}
\label{eq:LTSPST}
 S^k_+ ({p}^\mu)=\mathnormal{Q}_+({\bf p}) \frac{\sigma^k}{2} ({k}^\mu) \mathnormal{Q}^{-1}_+({\bf p}),
\end{eqnarray}
where $\mathnormal{Q}_+({\bf p})=\exp{\boldsymbol{\sigma}\cdot{\boldsymbol{\zeta}}}/2$. 
We will confirm this fact by using direct calculation in this section.

 Prior to manipulate the right-hand side of Eq. (\ref{eq:spin}),
 let us define
 $\cosh \frac{\zeta}{2} = \sqrt{\frac{p^0+m}{2m}}$ and $\sinh \frac{\zeta}{2} = \sqrt{\frac{p^0-m}{2m}}$
 with
 $(p^0)^2 = |\mathbf{p}|^2 + m^2$.
 One can then manipulate the right-hand side of  $S^k_+({p}^\mu)$ in Eq. (\ref{eq:spin}) such as
\begin{subequations}
 \begin{eqnarray}
  S^k_+ ({p}^\mu)
  &=& \frac{\sigma^k}{2} + \sinh \zeta   \ A^k
        + (\cosh \zeta -1)  \ B^k
 \\
 &=& \frac{\sigma^k}{2} + \sum_{n=1}
    \left[ \frac{\zeta^{2n-1}}{(2n-1)!}    \ A^k
        + \frac{\zeta^{2n}}{2n!}   \ B^k
         \right],
  \label{eq:series}
 \end{eqnarray}
\end{subequations}
 where $A^k = i\left({\boldsymbol{\sigma}} \times \hat{\mathbf{p}}\right)^k/2$
 and $B^k = \sigma^k/2- \hat{p}^k \left({\boldsymbol{\sigma}} \cdot \hat{\mathbf{p}} \right)/2$ with $\hat{\boldsymbol{\zeta}}=\hat{\bf p}$.
 One can notice
 that Eq. (\ref{eq:series}) can be expressed as a form of
 $e^X Y e^{-X} = Y + \sum_{n=1} X_n/n!$ with $X_{n+1}=\frac{1}{n+1}[X, X_n]$ for $n\ge 1$ and $X_1=[X, Y]$ with $Y=\sigma^k/2$
 in the Baker-Hausdorff formula.
 Then, let us work out an explicit form of the operator $X$ by assuming
 the transformation operator as $U_+ = \exp[ X ]$,
 where $X=f({\boldsymbol{\sigma}}, {\bf p})$ is a function of the rest spin operator
 ${\boldsymbol{\sigma}}/2$ and the momentum ${\bf p}$.
 In terms of the function $f({\boldsymbol{\sigma}}, {\bf p})$, the recursive relation is given as
 $X_{n+1} = \frac{1}{n+1} [ f({\boldsymbol{\sigma}}, {\bf p}), X_{n}]$ with
 $X_1 = [ f({\boldsymbol{\sigma}}, {\bf p}), \sigma^k/2]$.
 Comparing with Eq. (\ref{eq:series}),
 we have the two relations
 $X_{2n-1} = \frac{\zeta^{2n-1}}{(2n-1)!}
                  \left({\boldsymbol{\sigma}}/2 \times \hat{\mathbf{p}}\right)^k $
 and
 $
 X_{2n} = \frac{\zeta^{2n}}{2n!}  \left( \sigma^k/2
        - \hat{p}^k \left({\boldsymbol{\sigma}}/2 \cdot \hat{\mathbf{p}}\right)\right).
 $
 In determining the function $f({\boldsymbol{\sigma}}, {\bf p})$, thus, we have the two conditions
 $X_1 = [ f({\boldsymbol{\sigma}}, {\bf p}), \sigma^k/2]
 = \zeta \left({\boldsymbol{\sigma}}/2 \times \hat{\mathbf{p}}\right)^k $
 and
 $   \frac{1}{2n+1} [ f({\boldsymbol{\sigma}}, {\bf p}), \left({\boldsymbol{\sigma}}/2 \times \hat{\mathbf{p}}\right)^k ]
  = \frac{\zeta}{2n}  \left( \sigma^k/2
        - \hat{p}^k \left({\boldsymbol{\sigma}}/2 \cdot \hat{\mathbf{p}}\right)\right).$
 By using $[\sigma^i, \sigma^j] = 2 i \epsilon_{ijk}\sigma^k$,
 we see
 $\left({\boldsymbol{\sigma}} \times \hat{\mathbf{p}}\right)^k
 = [ \sigma^j, \sigma^k] \hat{\mathbf{p}}^j/2 $
 and then find $f({\boldsymbol{\sigma}}, {\bf p}) =\zeta\, \sigma^j \hat{\mathbf{p}}^j/2=\boldsymbol{\sigma}\cdot{\bf p}/2$.
 By putting the function $f({\boldsymbol{\sigma}}, {\bf p}) $ into
 the second condition, one can find that the equality of the second condition holds.
 Finally, $S^k_+({p}^\mu)$ in Eq. (\ref{eq:spin}) is re-expressed as
\begin{equation}
  S^k_+({p}^\mu)
  =
  \exp\!\Big[\frac{1}{2} {\boldsymbol{\sigma}}\cdot {\boldsymbol{\zeta}}\Big]
  \left( \frac{\sigma^k}{2}  \right)
  \exp\!\Big[-\frac{1}{2} {\boldsymbol{\sigma}}\cdot {\boldsymbol{\zeta}} \Big].
 \nonumber
\end{equation}
 Consequently,
 the spin oprator $S^k_+({p}^\mu)$
 is described by the similarity transformation of the rest spin operator $S^k({k}^\mu)$ with 
\begin{equation}
 \mathnormal{Q}_+({\bf p})= \exp\left[ \frac{{\boldsymbol{\sigma}}}{2}\cdot {\boldsymbol{\zeta}} \right].
  \nonumber
\end{equation}
 Similarly, we also obtain
 $\mathnormal{Q}_-({\bf p})= \exp\left[- {\boldsymbol{\sigma}}/2\cdot {\boldsymbol{\zeta}} \right] $
 from $S^k_-({p}^\mu)$ in Eq. (\ref{eq:spin}).

%

 \underline{~\hspace*{4cm}}




\end{document}